\documentclass[twocolumn]{aastex631}
\usepackage{sln_commands}
\usepackage{meth_quantities}

\shorttitle{Discovery of 25 \micron{} Methanol}
\shortauthors{Nickerson et al.}

\begin{document}

\title{The Discovery of 25 \micron{} Interstellar Methanol}

\correspondingauthor{Sarah L. Nickerson}
\email{sarah.nickerson@nasa.gov}

\author[0000-0002-7489-3142]{Sarah L. Nickerson}
\affiliation{Space Science and Astrobiology Division, NASA Ames Research Center, Moffett Field, CA 94035, USA}
\affiliation{Bay Area Environmental Research Institute, Moffett Field, CA, 94035, USA}

\author[0000-0001-9920-7391]{Naseem Rangwala}
\affiliation{Space Science and Astrobiology Division, NASA Ames Research Center, Moffett Field, CA 94035, USA}

\author[0000-0002-8030-7410]{Keeyoon Sung}
\affiliation{Jet Propulsion Laboratory, California Institute of Technology, Pasadena, CA 91011, USA}

\author[0000-0003-2458-5050]{Xinchuan Huang}
\affiliation{Space Science and Astrobiology Division, NASA Ames Research Center, Moffett Field, CA 94035, USA}
\affiliation{SETI Institute, 339 Bernardo Ave, Suite \#200, Mountain View, CA 94043, USA}

\author[0000-0003-2553-4474]{Edward J. Montiel}
\affiliation{U.S. Naval Observatory, 3450 Massachusetts Avenue NW, Washington, DC 20392, USA}
\affiliation{SOFIA-USRA, NASA Ames Research Center, MS 232-12, Moffett Field, CA 94035, USA}

\author[0000-0002-6528-3836]{Curtis DeWitt}
\affiliation{Space Science Institute, 4765 Walnut St, Suite B, Boulder, CO 80301, USA}

\author[0000-0002-5714-799X]{Valentin J. M. Le Gouellec}
\affiliation{Institut de Ci\'encies de l'Espai (ICE-CSIC), Campus UAB, Can Magrans S/N, E-08193 Cerdanyola del Vall\`es, Catalonia, Spain}
\affiliation{Institut d’Estudis Espacials de Catalunya (IEEC), c/Gran Capita, 2-4, E-08034 Barcelona, Catalonia, Spain}
\affiliation{Space Science and Astrobiology Division, NASA Ames Research Center, Moffett Field, CA 94035 USA}

\author[0000-0001-6275-7437]{Sean W. J. Colgan}
\affiliation{Space Science and Astrobiology Division, NASA Ames Research Center, Moffett Field, CA 94035, USA}

\author[0000-0001-7730-2240]{Jason Dittmann}
\affiliation{Department of Astronomy, University of Florida, Gainesville, FL 32611, USA}

\author[0000-0002-6556-6692]{Jos\'e Pablo Fonfr\'ia}
\affiliation{Instituto de F\'isica Fundamental (IFF-CSIC), C/ Serrano 121, E-28006 Madrid, Spain}

\author[0000-0002-7042-4541]{Graham M. Harper}
\affiliation{Center for Astrophysics and Space Astronomy, University of Colorado, 389 UCB, Boulder, CO 80309, USA}

\author[0000-0002-2626-7155]{Kathleen E. Kraemer}
\affiliation{Institute for Scientific Research, Boston College, 140 Commonwealth Avenue, Chestnut Hill, MA 02467, USA}

\author[0000-0003-0665-6505]{Jialu Li}
\affiliation{Department of Astronomy, University of Maryland, College Park, MD 20742, USA}

\author[0000-0001-9540-9121]{Conor A. Nixon}
\affiliation{Planetary Systems Laboratory, NASA Goddard Space Flight Center, MS 693, 8800 Greenbelt Road, Greenbelt, MD 20771, USA}

\author[0000-0002-3936-2469]{Maisie Frances Rashman}
\affiliation{School of Physical Sciences, The Open University, Milton Keynes, UK}

\author[0000-0002-7853-6871]{Clara Sousa-Silva}
\affiliation{Physics, Bard College, 30 Campus Road, Annandale-on-Hudson, NY 12504, USA}
\affiliation{Institute of Astrophysics and Space Sciences, Rua das Estrelas, Porto, 4150-762, Portugal}

\author[0000-0003-0306-0028]{Alexander G. G. M. Tielens}
\affiliation{Department of Astronomy, University of Maryland, College Park, MD 20742, USA}

\author[0000-0002-9123-0068]{William D. Vacca}
\affiliation{NSF’s NOIRLab, 950 N. Cherry Avenue, Tucson, AZ 85719, USA}

%TC:ignore
\begin{abstract}

%todo check final numbers

We present the first astrophysical detection of methanol (\methanol{}) in the torsional band near 25 \micron{}. Using high resolution mid-infrared (MIR) spectroscopy, we identified over seventy gas-phase \methanol{} absorption lines between 20 and 28 \micron{} towards the massive protostar NGC 7538 IRS 1 with SOFIA/EXES. We derive a temperature of 180 K and a total column density of $2\times10^{17}$ cm$^{-2}$, comparable to sub-mm measurements. Complementary analysis of acetylene (\acet{}) absorption lines is also included. Both \methanol{} and \acet{} reveal an unresolved second velocity component. These MIR absorption lines likely probe the molecular material in two edge-on disks, supporting the scenario that NGC 7538 IRS 1 consists of multiple protostars.  We provide an updated line list for the torsional band of \methanol{}, which was generated from lab work and model calculations. This discovery and the updated line list will enable the search for \methanol{} in JWST/MIRI spectra. 

\end{abstract}
%TC:endignore

%todo keywords
% \keywords{Classical Novae (251) --- Ultraviolet astronomy(1736) --- History of astronomy(1868) --- Interdiskiplinary astronomy(804)}

\section{Introduction} \label{sec:introduction}

Methanol (\methanol{}) is prevalent in a wide range of star-forming environments, including in its solid phase towards molecular clouds \citep{McClure2023} and protostars \citep{Boogert2008,Chen2024}; and in its gas phase towards prestellar cores \citep{Soma2015,Punanova2022}, massive star-forming regions \citep{Blake1987}, photo-dissociation regions \citep{Cuadrado2017}, protostars \citep{VanDishoeck1995,Maret2005}, and protoplanetary disks \citep{Walsh2016}.

\methanol{} is a key stepping-stone to the formation of more complex organic molecules in protostellar environments \citep{Charnley1992,Bennett2007,Garrod2008,Oberg2009,Aikawa2020,Jin2020}, including  precursors of prebiotic molecules \citep{Boyer2016,Catone2021}. Gas-phase reactions produce an insufficient amount of \methanol{} to explain its high abundances \citep{Garrod2006,Turner1998,Guzman2013}.  Laboratory experiments and observations agree that \methanol{} most easily forms in cold, dense environments in the interstellar medium (ISM) by the successive hydrogenation of CO catalyzed on the surface of icy dust grains \citep{VanDerTak2000,Watanabe2002,Fuchs2009,An2017,Zhao2023}.

It is possible that the molecules formed in cold, dense pre-stellar clouds will survive to be incorporated into planetesimals and planets, influencing the formation of life \citep{Sandford2020}. This is supported by recent evidence that the \methanol{} discovered at sub-mm wavelengths around protoplanetary disks was directly inherited from the cold, dark cloud phase \citep{Booth2021,Booth2023}. The molecules around protoplanetary disks subsequently feed the chemical composition of  planets \citep{Ilee2021,Oberg2021}.

The bridge linking protoplanetary disks, low mass protostars, and the cold clouds is the ``hot corino'' phase. These small regions of warm, dense molecular gas are the result of evaporation from icy grain mantles heated by the newly-formed protostars \citep{Ceccarelli2004,Beltran2018}. Around massive protostars, these regions are referred to as ``hot cores'' \citep{VanDishoeck1998,Kurtz2000,VanDerTak2004,Cesaroni2005}, which serve as a powerful probe of the chemical composition of the interstellar medium.
%, and given that our own Sun may have formed in a high mass star-forming region \citep{Adams2010}, it is possible that our planetary system inherited its molecular inventory from a previous hot core phase.

\begin{figure}

\centering
\includegraphics[height=0.29\textheight]{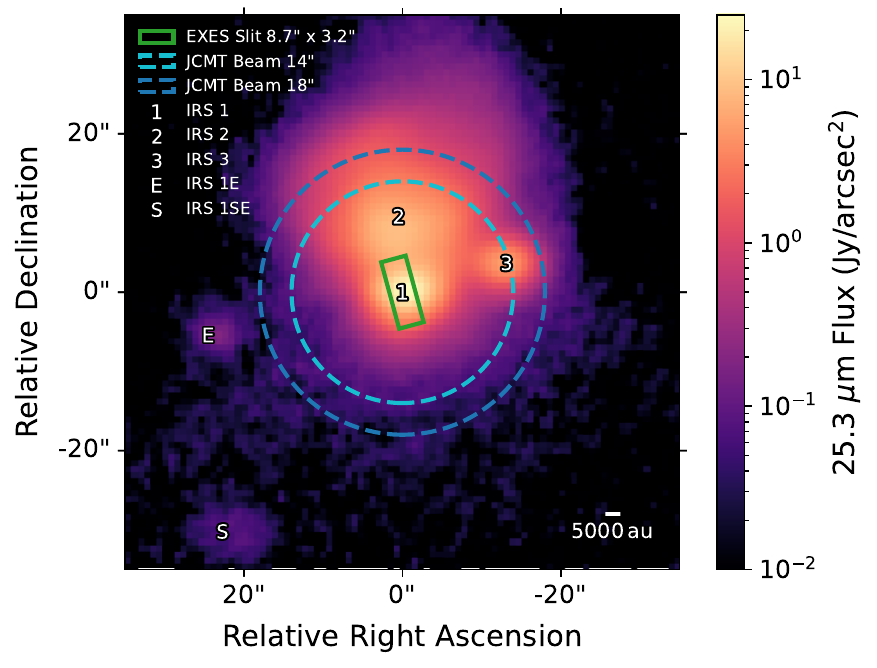}
\caption{25.3 \micron\ map of the NGC 7538 region from SOFIA/Faint Object infraRed Camera (FORCAST; \citealt{Herter2013}) archival data (Cycle 1 Program 0034, PI A. G. G. M. Tielens). Our target and brightest source, IRS 1, is at the centre with offset 0\arcsec{}, 0\arcsec{} corresponding to $\alpha$(J2000) = 23:13:45.37, $\delta$(J2000) = +61:28:10.5. The positions of IRS 1 and the dimmer sources (IRS 2, IRS 3, IRS 1E, and IRS 1SE) are given by \citet{Sandell2020}. In solid green is the 8.7\arcsec{}$\times$3.2\arcsec{} EXES slit from this work's 23.9 \micron{} setting (Table \ref{tab:obsexes}), and in dashed blue the 14\arcsec{} and 18\arcsec{} JCMT beams \citep{VanDerTak2000,Bisschop2007} whose \methanol{} results we will use for a comparison (Section \ref{sec:prevobs}). au scale uses a distance of 2.65 kpc \citep{Moscadelli2009}. Note the logarithmic scale for the flux, which exaggerates the brightness of the non-IRS 1 sources. \label{fig:bigmap}}
\end{figure}

%IRS 1, first detected as an OH maser source \citep{Hardebeck1971},
One such hot core surrounds the massive protostar NGC 7538 IRS 1 (Figure \ref{fig:bigmap}), which is the brightest infrared \citep{Wynn-Williams1974} and radio \citep{Martin1973} source within a molecular cloud in the \hii{} region NGC 7538 \citep{Sharpless1959,Habing1972,Werner1979}. With a mass of 30--50 \msun{} \citep{Willner1976,Hackwell1982,Sandell2020},
IRS 1 drives a large-scale, bipolar molecular outflow in the region \citep{Campbell1984,Scoville1986,Kameya1989,Davis1998,Sandell2020}. \methanol{} masers observed at mm and radio frequencies provide evidence for a circumstellar, edge-on disk around the central protostar \citep{Minier2000,Wiesemeyer2004,Pestalozzi2004} and a possible torus \citep{Surcis2011}.  \citet{Sandell2020} demonstrated that IRS 1's formation process was similar to that of low-mass stars, fed by an accretion disk that drives the outflow, but in a denser, higher-accreting environment. The disk may be hidden by a strong, ionised jet. It is debated whether NGC 7538 IRS 1 is a single \citep{Sandell2020}, binary \citep{Beuther2017}, or triple system \citep{Moscadelli2014,Goddi2015,Moscadelli2025}.
%Its luminosity ranges from $\sim 2\times10^4$ to $>10^5$ \lsun{}, giving it a spectra type of O6 to O8.5 and a mass of 30--50 \msun{} \citep{Willner1976,Hackwell1982,Sandell2020}. It is in the Perseus Arm at a distance of 2.65 kpc  \citep{Moscadelli2009}.

High mass star-forming regions, such as NGC 7538 IRS 1, have been the subject of numerous high spectral resolution ($R\sim$ 60,000 to 100,000) infrared observations that have led to the measurement of the individual transitions of molecular species \citep{Knez2009,Indriolo2015,Barr2018,Rangwala2018,Dungee2018,Goto2019,Barr2020,Nickerson2021,Indriolo2020,Li2022,Barr2022,Barr2022a,Nickerson2023,Li2023}. The spectral region around 25 \micron{} has been less explored in these MIR surveys compared to shorter wavelengths. High resolution is crucial for the secure identification of new molecules \citep[e.g. HNC,][]{Nickerson2021}. 

In this Letter, we announce the discovery of the torsional band of \methanol{} at 25 \micron{}, observed towards the massive protostar NGC 7538 IRS 1 in high resolution spectra taken by the Echelon-Cross-Echelle Spectrograph instrument (EXES; \citealt{Richter2018}) aboard the Stratospheric Observatory For Infrared Astronomy (SOFIA; \citealt{Young2012}). We also provide a complementary analysis of acetylene (\acet{}) towards NGC 7538 IRS 1 to contextualize the \methanol{} measurements. In Section \ref{sec:observations} we describe our observations with EXES, in Section \ref{sec:methods} we share our methods, in Section \ref{sec:results} we show the results, in Section \ref{sec:discussion} we discuss the implications of our findings, and we summarize our conclusions in Section \ref{sec:conclusions}.

\section{Observations}\label{sec:observations}

Our observations were carried out with SOFIA/EXES between May 3 and 6, 2022 as part of the ``The SOFIA/EXES Mid-IR High Spectral Resolution Library'' Legacy Survey (PI E. J. Montiel, Program ID 75\_0106). All observations were in the high-low mode, with a slit width of 3.2\arcsec. We detected \methanol{} between the 19.7 and 28 \micron{} settings.

We also complement our methanol spectrum with analysis of the \acet{} spectrum observed in the 13.9 \micron{} setting of the Legacy Survey, and with SOFIA archival data from the NASA/IPAC Infrared Science Archive (IRSA)\footnote{\url{https://irsa.ipac.caltech.edu/}} at 7.8 \micron{} (PI A. G. G. M. Tielens, Cycle 09 Program 0072) and 7.3 to 7.6 \micron{} (PI H. Yorke, Program ID 76\_0004). 

\begin{figure*}
\centering
\includegraphics[width=0.99\textwidth]{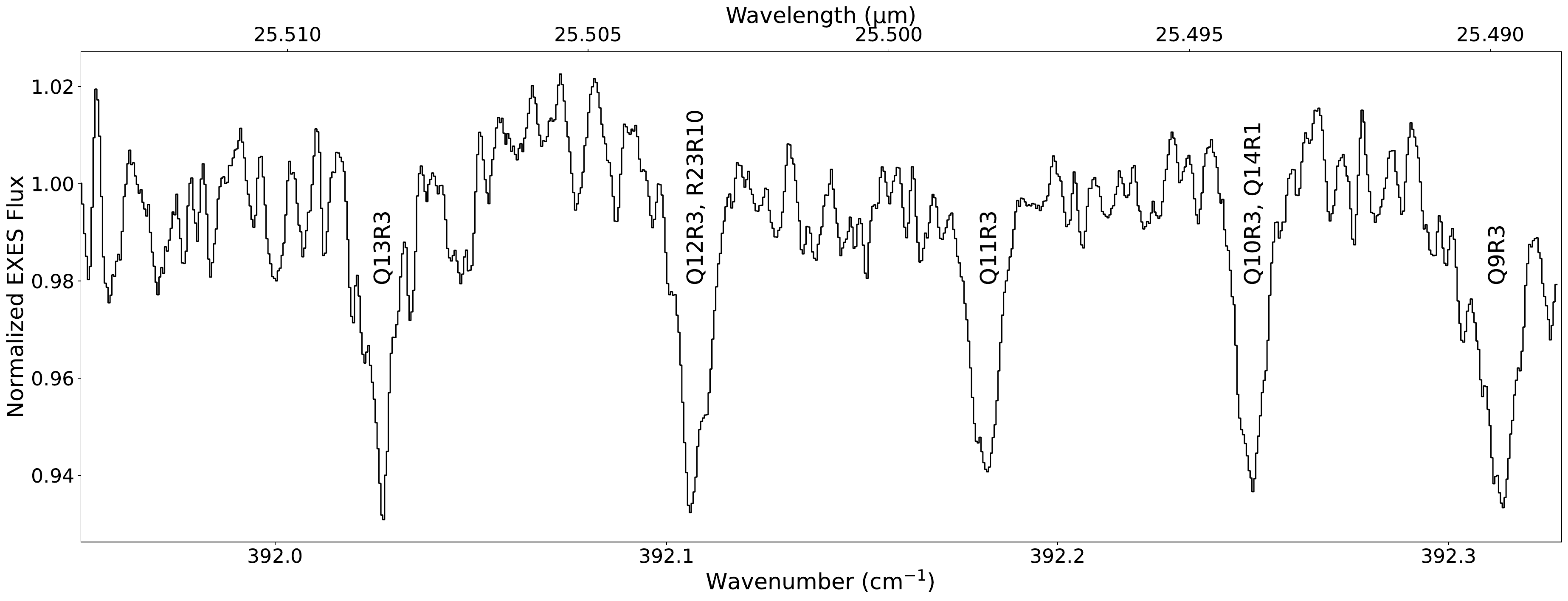}
\caption{Several \methanol{} lines in normalized flux as observed by EXES towards NGC 7538 IRS 1. Transition labels are indicated, all of which are 2$\nu_{12}$ and A-type, except Q14R1, which is E-type. Note the two blended lines. There are no telluric lines present in this plot. Other features are instrumental noise.} \label{fig:lines}
\end{figure*}

Figure \ref{fig:lines} presents an example \methanol{} spectra. Table \ref{tab:obsexes} and Figure \ref{fig:obsexes} summarize our observations (Appendix \ref{ap:obs}). All observations had resolving power $R\sim 60,000$. The total wavelength calibration errors are 0.5--1 \kms{} \citep{DeWitt2023}.

\section{Methods}\label{sec:methods}

Our analysis methods and equations are detailed in \citet{Nickerson2021,Nickerson2023}. The EXES flux is normalized and the atmosphere is divided out using a model from ATRAN \citep{Lord1992} and smoothed to the resolution of each observation. We then fit each individual absorption line to a Gaussian profile to find $\tau_0$ the line centre optical depth, \vlsr{} the line's central velocity in the local standard of rest (LSR) frame, and \vfwhm{} the full-width half maximum. We estimate $N_l$, the transition column density, as function of $\tau_0$ and \vfwhm{}.

We linearly fit the Boltzmann equation \citep{Goldsmith1999} to obtain the total column density, $N$, for a molecular species, and rotational temperature $T$. This assumes that the level populations of molecules are in local thermodynamic equilibrium, LTE.

We require $\lambda$ the rest wavelength of the transition, $A$ the Einstein coefficient for spontaneous emission, $g_{l}$ the lower statistical weight, $g_{u}$ the upper statistical weight, $Q(T)$ the partition function, and $E_{l}$ the energy level of the lower state for each molecular transition to complete the aforementioned calculations. For the torsional band of 25 \micron{} \methanol{}, we use an updated version of the line list from \citet{Brauer2012}, described in Appendix \ref{ap:methll}, which will be released in HITRAN2024. We use the partition functions of \citet{Villanueva2012} for each \methanol{} state ($A$- or $E$-type). For \acet{} we use HITRAN \citep{Hillman1991,Kabbadj1991,Gordon2022}.

\section{Results}\label{sec:results}

\begin{figure*}
\centering
\begin{tabular}{lcr}
\includegraphics[scale=0.4]{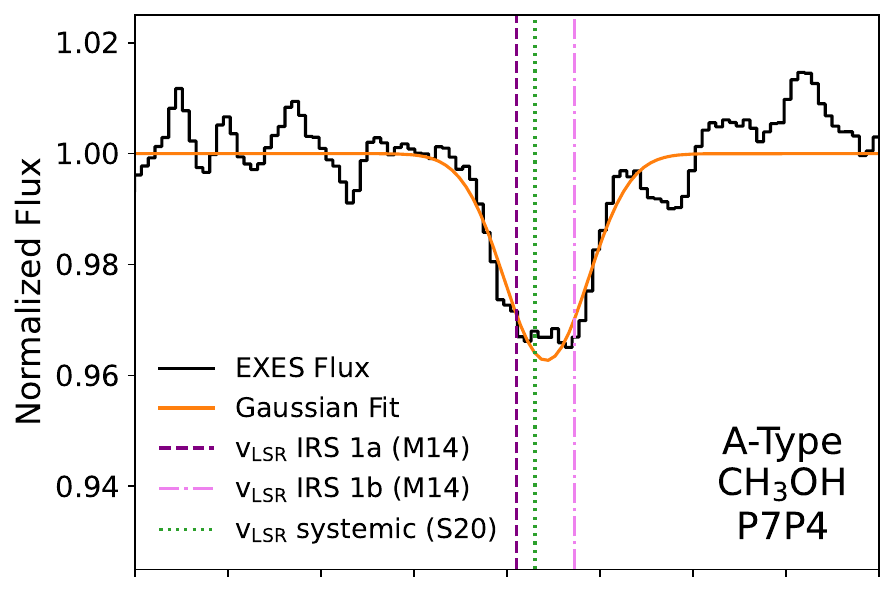}&
\includegraphics[scale=0.4]{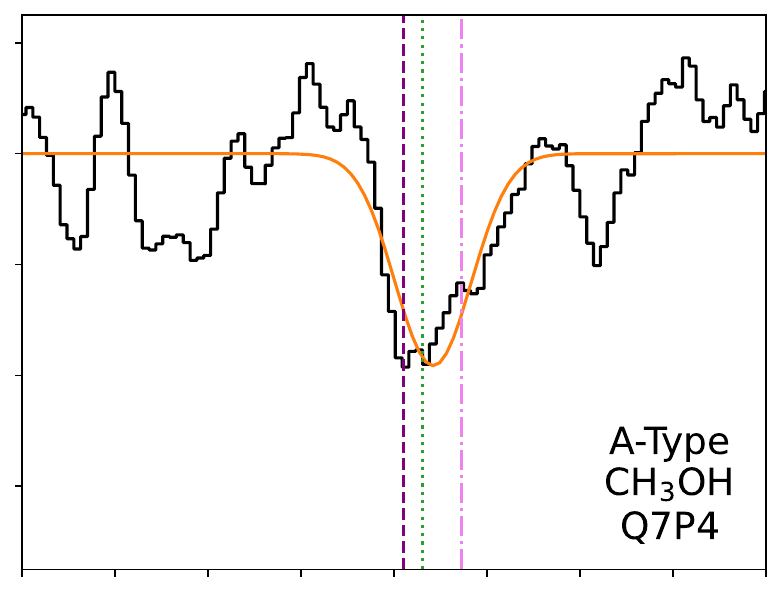}&
\includegraphics[scale=0.4]{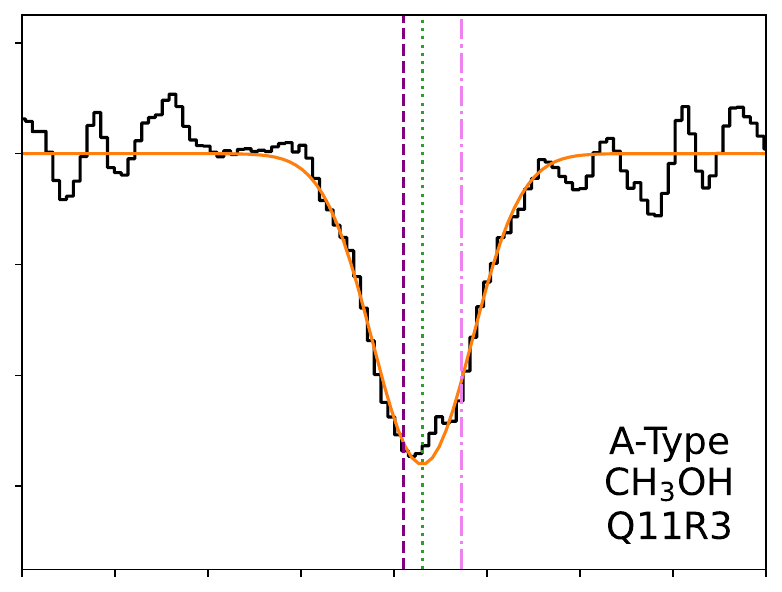}\\
\includegraphics[scale=0.4]{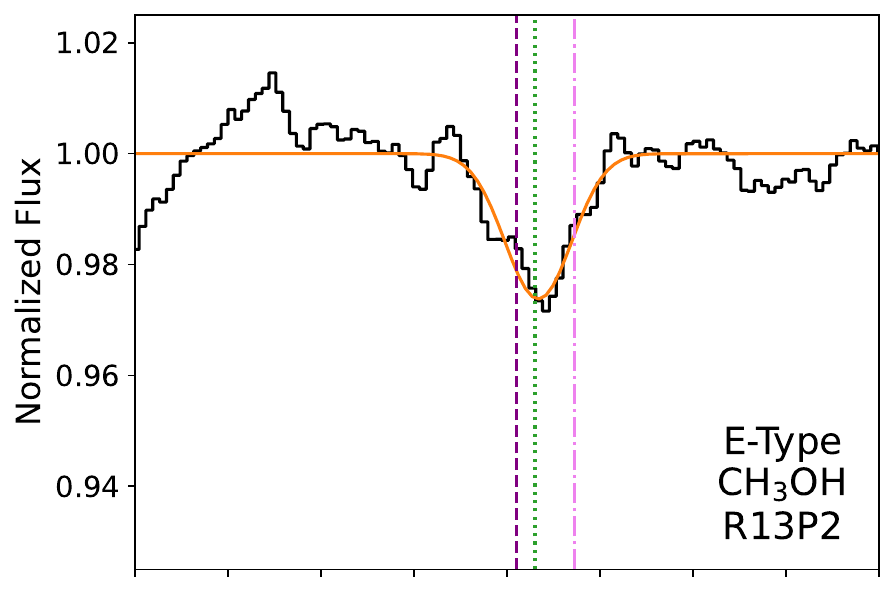}&
\includegraphics[scale=0.4]{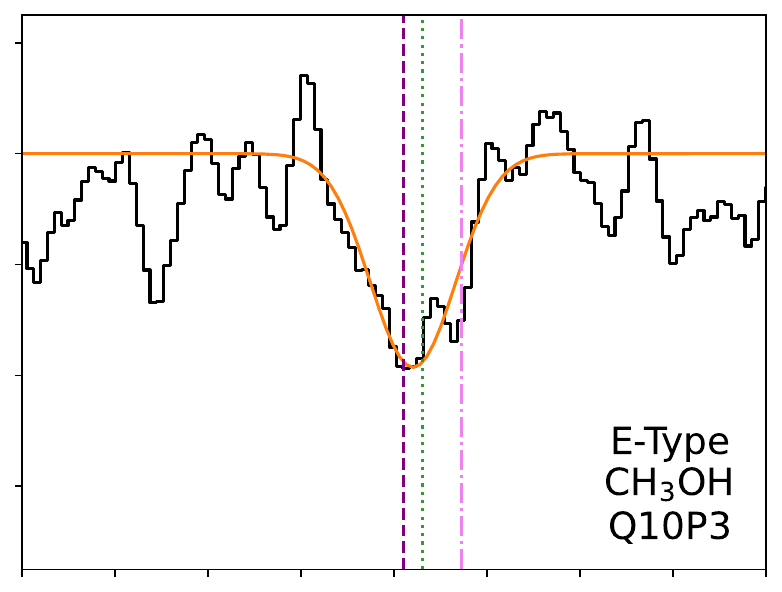}&
\includegraphics[scale=0.4]{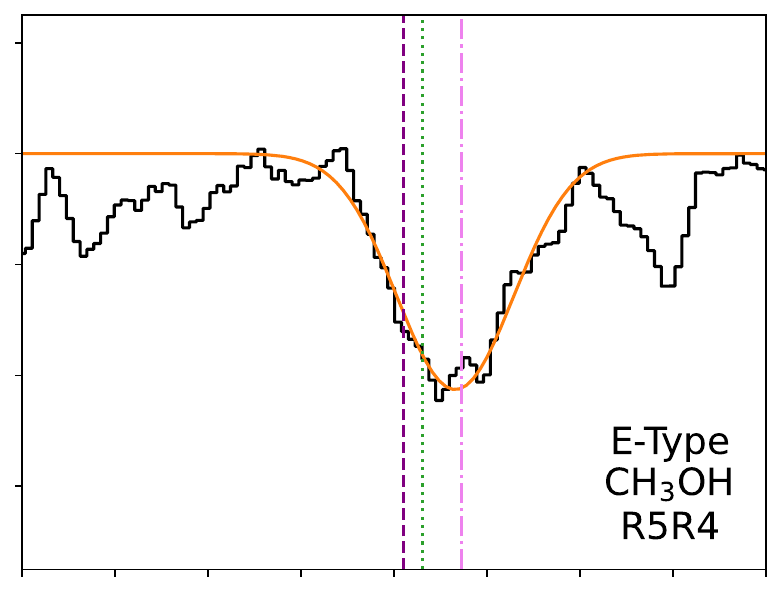}\\
\includegraphics[scale=0.4]{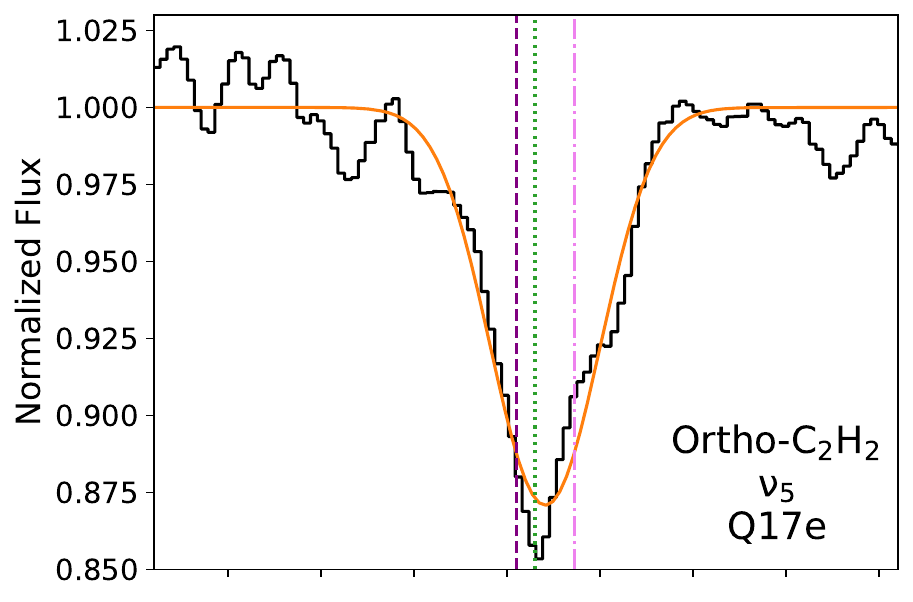}&
\includegraphics[scale=0.4]{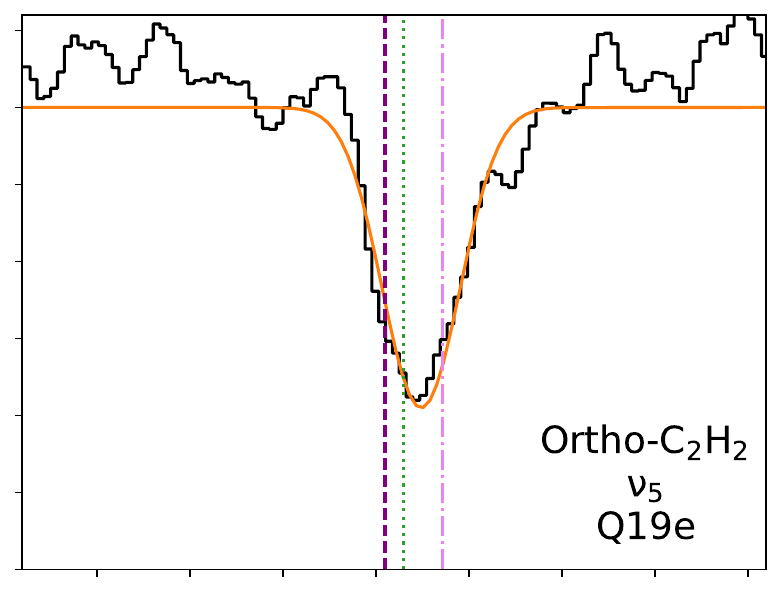}&
\includegraphics[scale=0.4]{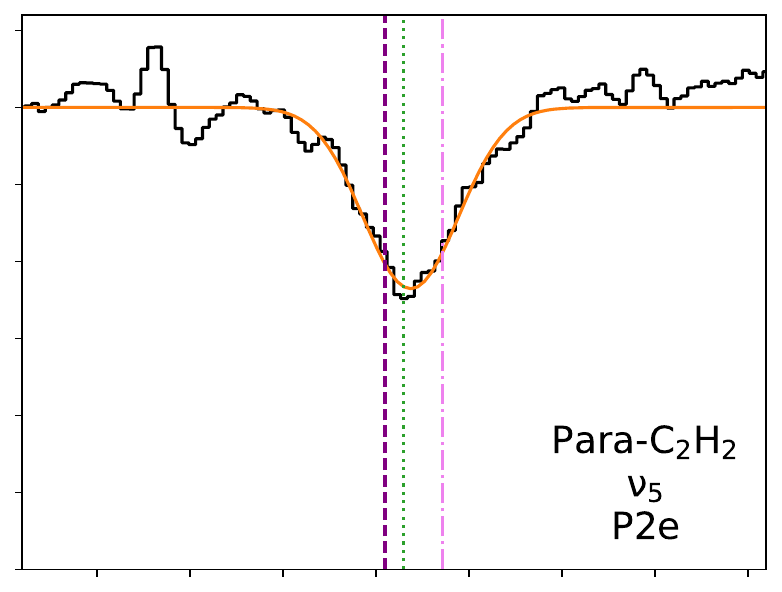}\\
\includegraphics[scale=0.4]{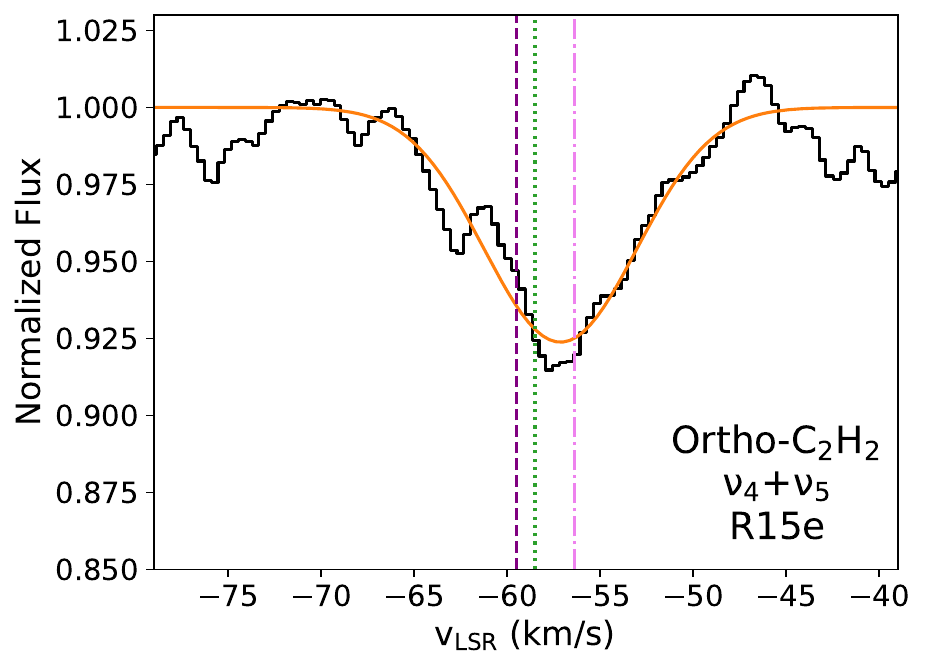}&
\includegraphics[scale=0.4]{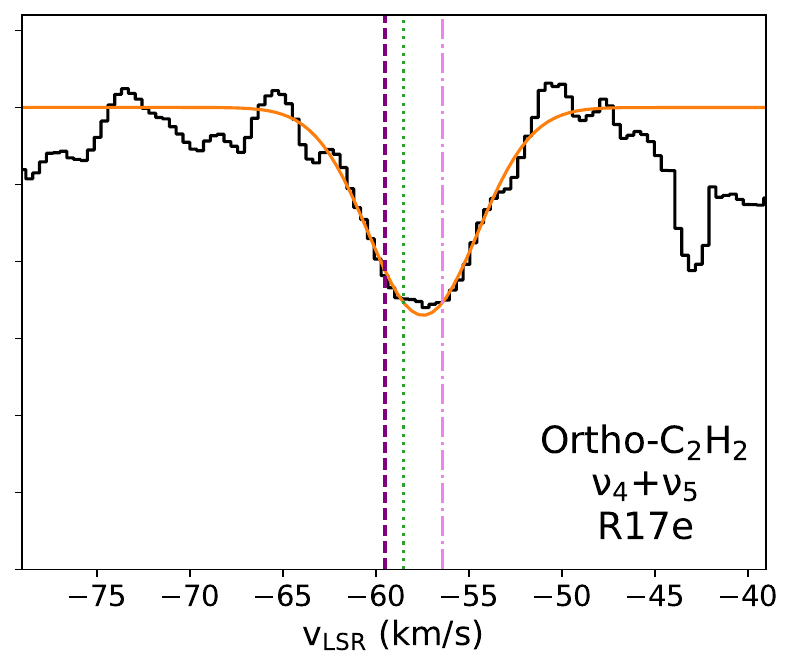}&
\includegraphics[scale=0.4]{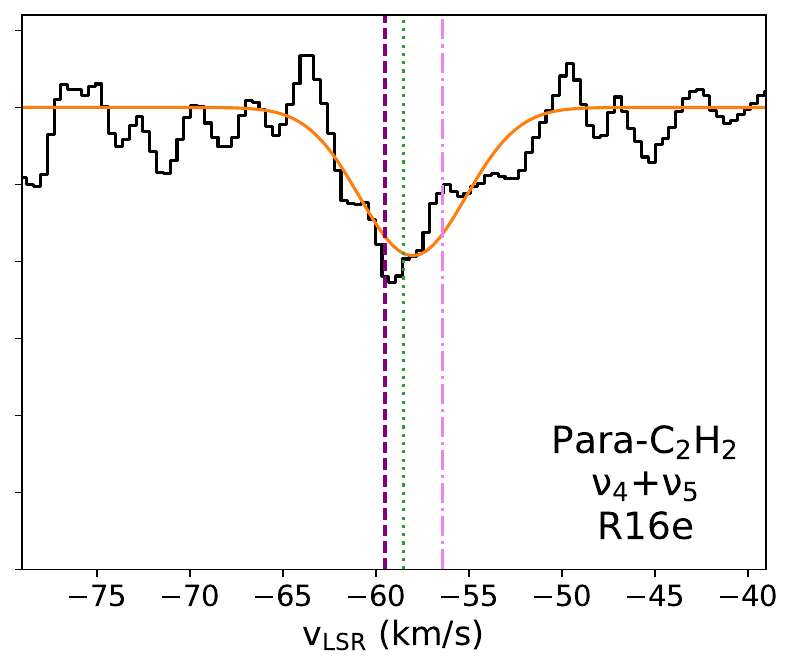}\\
\end{tabular}
\caption{Sample of Gaussian fits (solid orange) to molecular lines in normalized EXES spectra (solid black) towards NGC 7538 IRS 1 for A-type \methanol{} (top row), E-type \methanol{} (second row), the $\nu_5$ band of \acet{} (third row), and the $\nu_4+\nu_5$ band of \acet{} (bottom row). All \methanol{} lines displayed belong to the 2$\nu_{12}$ band. We have overlaid the \vlsr{} systemic to IRS 1 \citep[dotted green, $-58.5$ \kms{},][]{Sandell2020}, and those of the binary protostars IRS 1a and IRS 1b \citep[dashed purple, $-59.50$; and dash-dotted pink, $-56.39$ \kms{};][]{Moscadelli2014}. Note the different y-axis scales for \methanol{} and \acet{}.\label{fig:gauss}}
\end{figure*}

A sample of the single Gaussian fits to \methanol{} and \acet{} lines are shown in Figure \ref{fig:gauss}. The observed transitions and inferred parameters are given in Table \ref{tab:abslines} and a list of blended \methanol{} lines in Table \ref{tab:blendlines} in Appendix \ref{ap:lines}.

We see an asymmetry in the profiles of the deeper \acet{} and a few \methanol{} lines (see for example \acet{} Q17e in Figure \ref{fig:gauss}), revealing a second velocity component. However, double Gaussian fits to these lines were unsuccessful because the two components are too close for our observations' resolution.

\begin{figure*}
\centering
\gridline{\fig{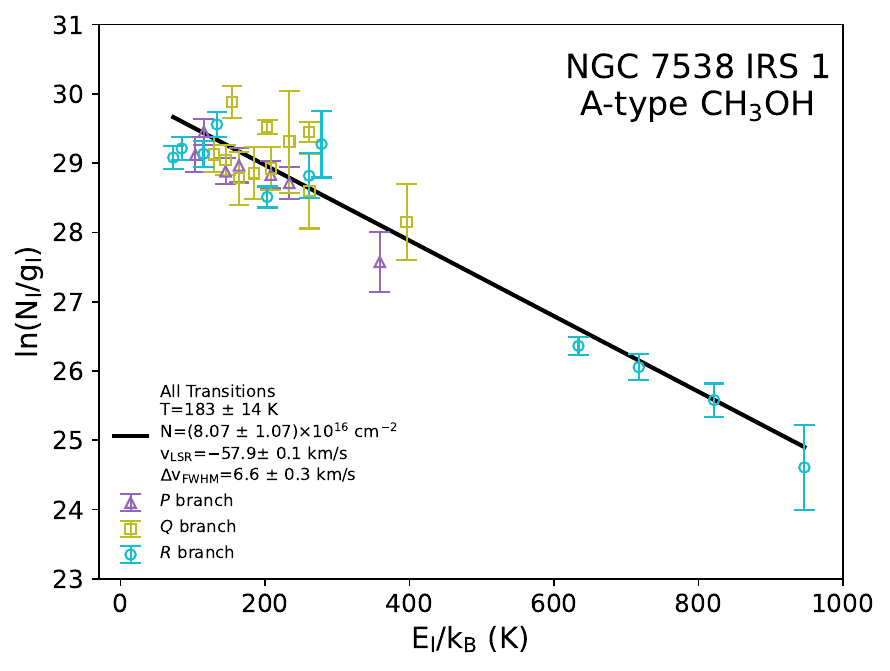}{0.49\textwidth}{}\hspace{-5mm}
          \fig{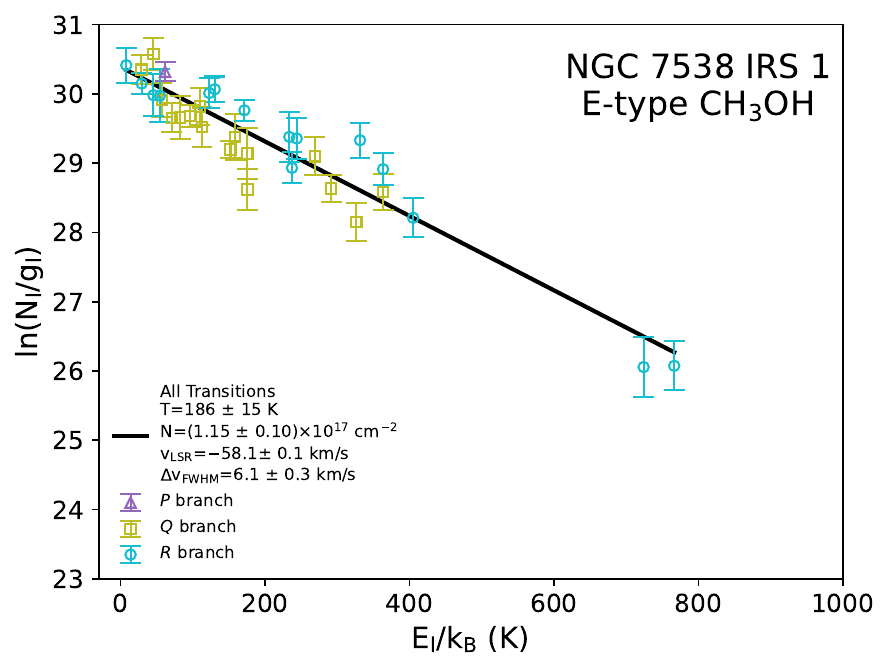}{0.49\textwidth}{}}
\vspace{-10mm}
\gridline{\fig{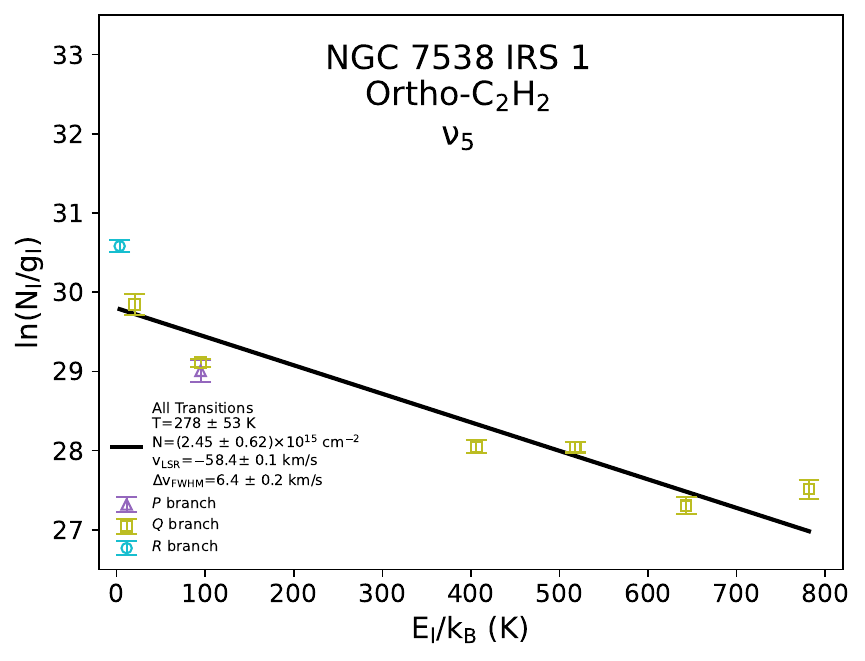}{0.49\textwidth}{}\hspace{-5mm}
          \fig{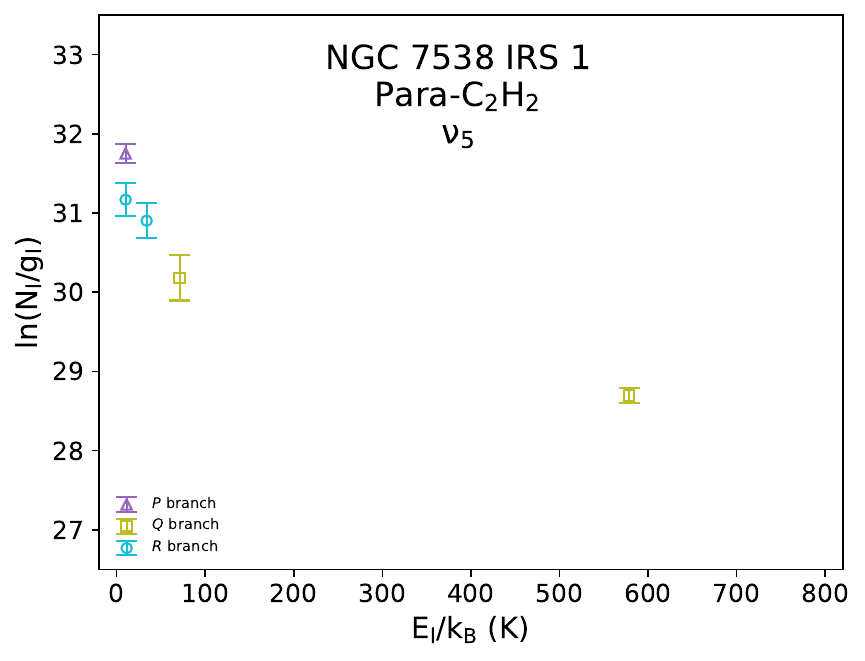}{0.49\textwidth}{}}
\vspace{-10mm}
\gridline{\fig{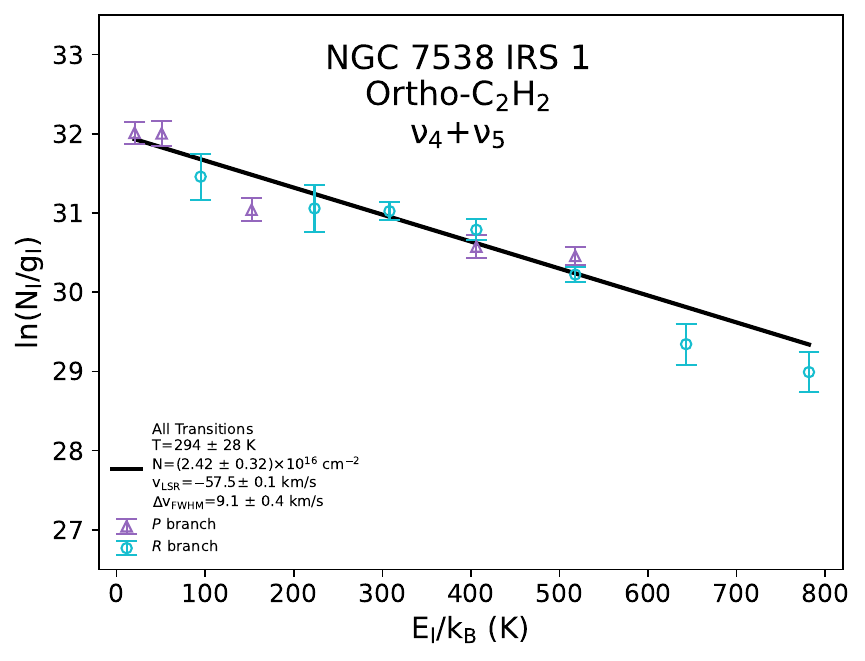}{0.49\textwidth}{}\hspace{-5mm}
          \fig{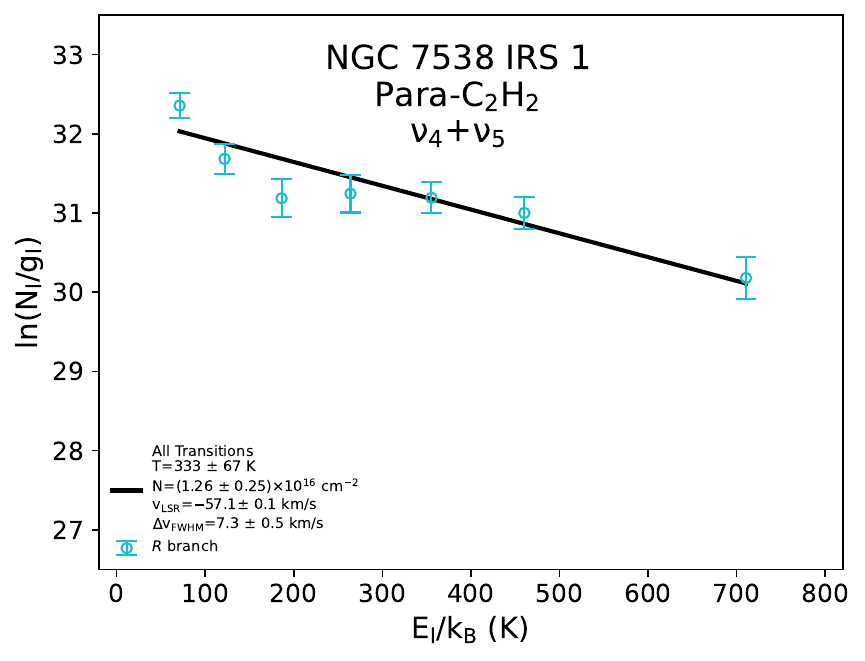}{0.49\textwidth}{}}          
\caption{Rotation diagrams for $A$-type (top left) and $E$-type (top right) \methanol{}; ortho- (centre left) and para-  (centre right) states of the $\nu_5$ band of \acet{}; and ortho- (bottom left) and para- (bottom right) states of the $\nu_{4}+\nu_{5}$ band of \acet{}. The $\nu_{12}$ band \methanol{} are the transitions with $E_l/k_B > 600$ K and $2\nu_{12}$ band \methanol{} with $E_l/k_B < 500$ K. The $P$, $Q$, and $R$ branch transitions are indicated with purple triangles, yellow squares, and blue circles respectively. \label{fig:rotmeth}}
\end{figure*}

The rotation diagram analysis is given in Figure \ref{fig:rotmeth} and the results are summarized in Table \ref{tab:rot}, alongside a comparison to previous publications. We ran curve of growth analysis on both species (Appendix \ref{ap:cog}) to verify the assumption that these lines are optically thin.

%todo verify final numbers
%todo double check other works' numbers
\begin{deluxetable*}{lllrrrrrrr}
\tablecaption{Overview of Species Properties towards NGC 7538 IRS 1\label{tab:rot}}
\tablehead{
\colhead{Ref} & \colhead{Beam Size} & \colhead{Type} & \colhead{Grouping}&\colhead{$\lambda$/$\nu$} &\colhead{\#}&\colhead{\vlsr} & \colhead{\vfwhm} &  \colhead{$T$} &
\colhead{$N$} \\
\colhead{} & \colhead{(\arcsec)} & \colhead{} & \colhead{} & \colhead{} &  \colhead{} & \colhead{(\kms)} & \colhead{(\kms)} & \colhead{(K)} &\colhead{(\csi)}}
\startdata
\hline
\multicolumn{10}{c}{\methanol{}}\\
This Work& 3.2$\times$5.4--12.6  &abs& A-type; $\nu_{12}$, $2\nu_{12}$& 25 \micron{}&29&--57.9$\,\pm\,$0.1&6.6$\,\pm\,$0.3$^{\dagger}$&183$\,\pm\,$14&(8.07$\,\pm\,$1.07)$\times 10^{16}$\\
& &abs&E-type; $\nu_{12}$, $2\nu_{12}$& 25 \micron{} &32&--58.1$\,\pm\,$0.1&6.1$\,\pm\,$0.3$^{\dagger}$&186$\,\pm\,$15&(1.15$\,\pm\,$0.10)$\times 10^{17}$\\
vdT00&14, 18&emi&$J = 5 \rightarrow$ 4,   7$\rightarrow$6&290 GHz&28&$-57.62$&3.9&189&$2.2\times10^{15}$\\
B07&14&emi&$J=7\rightarrow 6$&338 GHz&55&---&---&$156\pm10$&$1.2\times 10^{17}$\\
\hline
\multicolumn{10}{c}{\acet{}}\\
This Work& 3.2$\times$2.4--3.4  &abs&ortho, $\nu_5$&13.5 \micron{}&8&--58.4$\,\pm\,$0.1&6.4$\,\pm\,$0.2$^{\dagger}$&278$\,\pm\,$53&(2.45$\,\pm\,$0.62)$\times 10^{15}$\\
& &abs&para, $\nu_5$&13.5 \micron{}&5&--57.9$\,\pm\,$0.5&7.2$\,\pm\,$2.1$^{\dagger}$&---&---\\
& &abs&ortho, $\nu_4+\nu_5$& 7.6 \micron{} &12&--57.5$\,\pm\,$0.1&9.1$\,\pm\,$0.4$^{\dagger}$&294$\,\pm\,$28&(2.42$\,\pm\,$0.32)$\times 10^{16}$\\
& &abs&para, $\nu_4+\nu_5$& 7.6 \micron{} &7&--57.1$\,\pm\,$0.1&7.3$\,\pm\,$0.5$^{\dagger}$&333$\,\pm\,$67&(1.26$\,\pm\,$0.25)$\times 10^{16}$\\
K09&1.5$\times$8.62--10.61 &abs&C1; $\nu_5$, $\nu_4+\nu_5$,&13.5, 8,&&&&\\
&&&$\nu_4+\nu_5-\nu_4$&13.6 \micron{}&47&$-55.7\,\pm\,0.3$&$1.0\pm0.2$&$225\pm20$&$(3.0\,\pm\,0.6)\times 10^{16}$\\
&&abs&C2; $\nu_5$,$\nu_4+\nu_5$,&13.5, 8,&&&&\\
&&&$\nu_4+\nu_5-\nu_4$&13.6 \micron{}&47&$-59.4\,\pm\,0.3$&$1.0\pm0.2$&$191\pm10$&$(2.8\,\pm\,0.2)\times 10^{16}$\\
\hline
\enddata
\tablecomments{For each species we give: the reference, beam size, lines type (abs)orption or (emi)ssion, grouping, approximate central wavelength/frequency, the number of detected lines, central local standard rest of velocity, full-width half-maximum, the temperature, and the total column density. For this work and K09, we give the slit width$\times$range for slit length. Grouping is reference and species dependent. For this work the groupings are explained in Sections \ref{sec:resmeth} and \ref{sec:resacet} and we give the mean \vlsr{} and \vfwhm{}; in vdT00 and B07 the most prominent bands are given; in K09 three different bands of \acet{} are fit together in two components, C1 and C2, and ortho- and para-states are combined. References: vdT00 \citep{VanDerTak2000}, B07 \citep{Bisschop2007}, and K09 \citep{Knez2009}. $^{\dagger}$We give the observed \vfwhm{} (See Appendix \ref{ap:lines}). After allowing for instrumental resolution, the A- and E-type \methanol{} have intrinsic \vfwhm{}=4.3 and 3.5 \kms{}; $\nu_5$ o- and p-\acet{} have intrinsic \vfwhm{}=4.0 and 5.2 \kms{}; and $\nu_4+\nu_5$ o- and p-\acet{} have intrinsic \vfwhm{}=7.6 and 5.3 \kms{}.}
\end{deluxetable*}

\subsection{Methanol}\label{sec:resmeth}

\methanol{} occurs in two separate states determined by the spin alignment of the hydrogen atoms in the methyl group (CH$_3$): A- and E-type\footnote{A-type corresponds to parallel spins and a spin angular momentum of 3/2, while E-type has asymmetric spins and a spin angular momentum of 1/2.}. Because transitions between the two types are forbidden, we have analyzed them as separate species. We fit \methnuma{} A-type and \methnume{} E-type transitions and calculated their total column densities and temperatures separately. We also observe two bands: the fundamental torsional band $\nu_{12}$ and first overtone torsional band $2\nu_{12}$, which appear to be in local thermal equilibrium (LTE) and are fit on the same rotation diagram for each state. 

For A-type \methanol{} we find a column density of \methcoldena{} and a temperature of \methtempa{}, while for E-type \methanol{} we estimate \methcoldene{} and \methtempe{}. \methanol{} has an observed \vfwhm{} = \methvfwhmavg{} and an average \vlsr{} = \methvlsravg{}, similar to the systemic velocity of NGC 7538 IRS 1 \citep[$-58.5$ \kms{},][]{Sandell2020}. We measure an E/A ratio of \ea, which will be discussed in Section \ref{sec:earatio}. Using the column density \nhtwo{}$=4.9\times10^{22}$ \csi{} of warm gas \citep[259 K,][]{Goto2015}, we measure  a \methanol{} abundance of $\sim 4.0\times10^{-6}$.

\subsection{Acetylene}\label{sec:resacet}

%toco check final numbers/ratio
We separated \acet{} into ortho and para states, and measured two bands: $\nu_5$ and $\nu_4+\nu_5$. The $\nu_5$ band of para-\acet{} did not have enough transitions over a wide range of energies for an accurate fit. For the three other groups, we find an average temperature of $\sim 300$ K and, similar to \methanol{}, a \vlsr{} around $-58$ \kms. We measured column densities of \acetcoldenof{} for $\nu_5$ ortho-\acet{}, \acetcoldenoff{} for $\nu_4+\nu_5$ ortho-\acet{}, and \acetcoldenpff{} for $\nu_4+\nu_5$ para-\acet{}. This gives abundances of $5.00\times10^{-8}$, $4.94\times10^{-7}$, and $2.57\times10^{-7}$ respectively using the same value for \nhtwo{} as in Section \ref{sec:resmeth}. We will discuss the column density ratio between bands in Section \ref{sec:methorigin}.

\section{Discussion} \label{sec:discussion}

\subsection{Previous Methanol and Acetylene Spectroscopy Towards NGC 7538 IRS 1}\label{sec:prevobs}

In the MIR, NGC 7538 IRS 1 has been observed in low spectral resolution with space-based ISO-SWS \citep[$R\sim\ $1,500,][]{DeGraauw1996} in which individual molecular transitions were not resolved \citep{Lahuis2000,Boonman2003,Boonman2003a,Keane2001} and in high resolution with the ground-based IRTF/TEXES \citep[$R\sim\ $100,000,][]{Lacy2002} from 7.6 to 13.7 \micron{} \citep{Knez2009,Barentine2012}. 

The molecular absorption lines of multiple species measured by \citet{Knez2009} (Table \ref{tab:rot}) were resolved into two components C1 and C2, with \vlsr{} $=-55.7$ \kms{} and $-59.4$ \kms{} respectively for \acet{} (the difference a little more than the limit of TEXES's spectral resolution). As explained in Section \ref{sec:results}, we see two components in some lines for both species, but are only able to measure a single component for both \acet{} and \methanol{}, with \vlsr{} $\sim -58$ \kms{}. This suggests that our molecular lines largely originate in the C2 component, with some contribution from the C1 component. The lower spectral resolution of the EXES observations compared to TEXES was insufficient to resolve the second component.

Spectroscopic observations towards NGC 7538 IRS 1 at sub-mm wavelengths have detected \methanol{} \citep{VanDerTak2000,Bisschop2007}. This work's \methanol{} temperature and \vlsr{} are similar to the sub-mm measurements (Table \ref{tab:rot}), with the exception of \citet{VanDerTak2000}'s column density being much lower (the difference in column density between \citealt{VanDerTak2000} and \citealt{Bisschop2007} is likely because the latter corrects for beam dilution). The MIR absorbing gas is expected to originate from a region closer to the central protostar compared to the sub-mm-emitting gas (refer to Section \ref{sec:methorigin}). Absorption lines from EXES probe a pencil-beam the size of the source, while emission lines originate from a larger region corresponding to the size of the beam (Figure \ref{fig:bigmap}).

\subsection{The Origins of the Molecular Absorption Lines Towards NGC 7538 IRS 1}\label{sec:methorigin}

The topography of the innermost region of NGC 7538 IRS 1 is complex, giving rise to multiple interpretations. \citet{Sandell2020} find no evidence of binary stars with separations greater than 30 au in mm maps, and propose a single protostar with an inclined disk. On the other hand, kinematic signatures in absorbing \methanol{} gas reveal two protostars with parallel disks separated by $\sim$ 430 au, which are embedded in a common circumbinary envelope probed by NH$_3$ \citep{Beuther2017}. \citet{Moscadelli2014} found that \methanol{} maser clusters identify three embedded protostars, IRS 1a (\vlsr$=-59.50$ \kms{}), IRS 1b (\vlsr$=-56.39$ \kms{}), and IRS 1c (\vlsr$\sim-60$ \kms{}), the earlier two with edge-on disks, as supported by later \water{} maser observations \citep{Moscadelli2025}. All three of these protostars are in the centre of this work's EXES beam (located at the ``1'' with their separate positions indistinguishable on the scale of Figure \ref{fig:bigmap}). Radio mapping of NH$_3$ absorption lines and one \methanol{} line reveal a temperature of $\sim$ 280 K for the circumbinary envelope around IRS 1a and 1b, and $\sim$ 110 K for a second, southern core around IRS 1c \citep{Goddi2015}.

The \methanol{} temperature we estimate, $\sim180$ K, is above the 100 K temperature required to release \methanol{} from dust grains into the gas-phase \citep{VanDerTak2000}, suggesting that the \methanol{} has survived in the hotter gas. The location of the absorbing molecular gas that hosts the \methanol{}, however, is open to interpretation due to two unresolved issues: (1) the topography of NGC 7538 IRS 1, as explained above; and (2) the source of the MIR continuum emission towards massive protostars in general.

Some observational studies suggest that MIR continuum from massive protostars is dominated by the cavity walls \citep{DeBuizer2006,deWit2010,Zhang2013,Debuizer2017}, while \citet{Boley2013} determined that both the disk and cavity walls contribute. Modelling suggests that the MIR continuum originates internally to the disk, possibly via viscous heating \citep{Nazari2023}. From high resolution MIR images of NGC 7538 IRS 1, \citet{DeBuizer2005} propose that heated dust on the walls of outflow cavities are the source of the large-scale MIR emission parallel to the CO outflow, while a circumstellar disk produces the MIR emission perpendicular to this outflow.

We explore two scenarios for the origins of the \methanol{} gas in this work: in the disk(s), absorbing MIR continuum from even deeper within the disk(s); or the infall and outflow, absorbing MIR continuum from the cavity walls of the envelope. Of these two, the disk scenario is more likely.

\subsubsection{The Disk Scenario}

Previous high-resolution MIR spectroscopic studies towards massive protostars have favoured the scenario that the disks host the absorbing  molecular gas \citep[][]{Knez2009, Barr2020,Barr2022,Li2023}.

As discussed in Section \ref{sec:prevobs}, we see evidence for two unresolved components that likely correspond to the \citet{Knez2009} C1 and C2 components. They argued that both components of the absorbing molecular gas lie in an edge-on disk. Their work, however, predates the discovery that IRS 1 may consist of multiple protostars. Across all molecules measured by \citet{Knez2009}, they found \vlsr$=-56.4 \pm 0.3$ for C1 and and $-59.7 \pm 0.3$ \kms{} for C2. These fall within the errors of the velocities for the edge-on disks around IRS 1a and 1b \citep{Moscadelli2014}. Each component may correspond to material in the edge-on disk around each protostar.
% There may be some contribution from IRS 1c, but only weakly since the depth of its absorption lines is about a third of those towards IRS 1a and 1b \citep[see Figure 2 in][]{Goddi2015}.

The differing column densities of the two \acet{} bands provide further hints of a disk-origin. \acet{} has been observed in high abundances towards protostars and multiple studies suggest that, similar to \methanol{}, it has been released from the ice-phase \citep{Lacy1989,Kaiser1998,Lahuis2000,Carr2008,Knez2009}. However, the formation mechanism for \acet{} on ice is unclear \citep{Moore1998}, and high temperature gas chemistry may play a role in \acet{} production in hot cores \citep{Bast2013}. 

We find a higher \acet{} temperature than \methanol{}, but the \vlsr{} is similar. While they belong to the same velocity component, \methanol{} probes a cooler, outer layer\footnote{The disk layers of a massive protostar are illustrated in Figure 13 of \citet{Barr2020}.} with a temperature similar to the equilibrium dust temperature at 25 \micron{}. As observed in two other massive protostars \citep{Barr2020}, we measure a gradient in \acet{} column density where the ortho-state at 13.5 \micron{} has an order of magnitude lower column density than the same state at 7.6 \micron{}. Since bands at the two wavelengths share the same lower ground state level, the derived column density of the ground level are expected to be the same. \citet{Barr2020,Barr2022} attribute this as evidence for a disk with a difference in continuum size at 7 and 13 \micron{} whereby the 13 \micron{} \acet{} lines are filled in by the continuum emission. 

\subsubsection{The Infall and Outflow Scenario}

\citet{Beuther2012,Beuther2013} suggest that the line-of-sight towards IRS 1 is through a cavity carved out by the outflows, to explain the escape of infrared radiation. They argue that internal disk(s) would be face-on, because the envelope of an an edge-on disk would extinguish the infrared radiation. They, along with \citet{Zhu2013}, measure the mm/sub-mm absorption lines of several molecular species, and interpret the red-shifted absorption lines to be in-falling gas, and the blue-shifted as outflowing gas. 

\citet{Beuther2012} observe two components of \methanol{} absorption at \vlsr{}$\sim -59$ and $-55$ \kms{}, similar to the velocities we and \citet{Knez2009} observe in the MIR absorbing gas, red- and blue-shifted of IRS 1's systemic velocity (Figure \ref{fig:gauss}). In this scenario the MIR continuum source would likely be the cavity walls rather than the internally heated disk.

However, any in-falling material would comprise of the cold, interstellar gas that has yet to be heated and release molecules into the gas-phase. The molecular emission lines in the outflow have widths $\sim 20$ \kms{} \citep{Sandell2020}, much wider than this work's MIR absorption lines (Table \ref{tab:rot}). Thus, so long as the envelope is optically thin, we consider the protostellar disk(s) to be the most likely origin of absorption lines.

\subsection{\textit{E/A} Ratio}\label{sec:earatio}

Upon \methanol{} formation at cold temperatures of 10 K, the $E/A$ abundance ratio is expected to be 0.69 and approaches 1 as the gas heats up \citep{Friberg1988,Wirstrom2011}. \citet{Zhao2023} measured this ratio towards 55 massive star-forming regions and found that 70\% have an E/A ratio less than 1, peaking at 0.6. \citet{Macdonald1996} and \citet{Mendoza2018} found similar ratios of $\sim$1 towards massive protostars. \citet{Wirstrom2011} estimated a ratio of $1.09\pm0.1$ towards NGC 7538 IRS 1, but this was calculated from a temperature determined by only three lines.  We find an E/A ratio of \ea, slightly higher than the equilibrium ratio of 1. This suggests the possibility that some astrophysical process may have altered the E/A ratio after \methanol{}'s formation on ice grains \citep{Wirstrom2011}, possibly shocks \citep{Humire2025}. 

\subsection{Enabling Methanol Observations with JWST/MIRI}\label{sec:jwst}

The medium resolution Mid-Infrared Instrument (MIRI, \citealt{Rieke2015,Wright2015,Wright2023}) onboard JWST has the capacity to access the inner $\sim$ 1--10 au of disks around both low- and high-mass protostars, including protoplanetary disks (the``planet-forming zone'', \citealt{Williams2011}), while instruments such as ALMA, can probe the outer regions \citep{Dullemond2007,Kamp2018,Bosman2022,Banzatti2023,Zhang2024}. Whether planets form first in-situ or later migrate inwards towards the inner region \citep{Armitage2024}, the molecules probed by the MIR have a profound influence on future planetary systems.

High spectral resolution is crucial for the secure identification of new molecules \citep[e.g. HNC,][]{Nickerson2021}. With MIRI ($R\sim$ 1500--4000, \citealt{Jones2023}), individual lines of several molecules are blended into broader, unresolved features.

A narrow \methanol{} gas-phase feature peaks at $\sim9.7$ \micron{}, dominated by the $\nu_8$ band \citep[HITRAN,][]{Gordon2022}. However, at MIRI's resolution, this feature overlaps with the S(3) \htwo{} transition at 9.665 \micron{}. Furthermore, protostellar flux in this region is depressed by absorbing silicate dust. Thus, the 25 \micron{} torsional band of \methanol{} is, in all likelihood, the only opportunity to observe gas-phase \methanol{} with JWST/MIRI.

The torsional band of \methanol{} enables measurements of this critical molecule in the inner, circumstellar material from which planets are formed and elucidate the origins of prebiotic material.

\section{Conclusions}\label{sec:conclusions}

%tood check final numbers
We report the first astronomical detection of the torsional band of \methanol{} at 25 \micron{}. In Appendix \ref{ap:methll}, we provide an updated line list for this band. With high-resolution SOFIA/EXES spectra we have identified over 70 \methanol{} lines in absorption towards the massive protostar(s) NGC 7538 IRS 1. The high abundance we measure reveals that the MIR is comparable to the sub-mm when accounting for this molecule's total inventory. 

We also analyzed \acet{} absorption lines, and measured a slightly higher temperature and a similar average \vlsr{} to \methanol{}, meaning that the two species are found in the same component.

Both \methanol{} and \acet{} show an unresolved, second velocity component. The absorbing molecules most likely reside in two circumstellar edge-on disks around two high mass protostars, IRS 1a and IRS 1b.

The \methanol{} band at 25 \micron{} discovered in this work is the best opportunity to study \methanol{} in the MIR with JWST. While \methanol{} has been observed previously at longer wavelengths, infrared absorption transitions are capable of probing the regions closest to the central protostars and the innermost $\sim$1--10 au of protoplanetary disks. Our discovery enables the search for this gas-phase complex organic molecule in the critical regions that will host future planetary systems.

%% Also note that the akcnowlodgment environment does not support long amounts of text. If you have a lot of people and institutions to acknowledge, do not use this command. Instead, create a new \section{Acknowledgments}.
%TC:ignore
\section{Acknowledgements}
% \begin{acknowledgments}
We would like to thank John Pearson of the Jet Propulsion Laboratory for providing us with his model calculations on the \methanol{} energy levels in the 25 \micron{} region; Kinsuk Acharyya, Partha Bera, and Ryan Fortenberry for the helpful chemistry discussions; Maja Marminge for the initial work on \acet{} during her summer internship; Uma Gorti for her feedback; Luca Moscadelli for providing the coordinates of IRS 1a and 1b; and the anonymous referees for their helpful suggestions. This work is based in part on observations made with the NASA/DLR Stratospheric Observatory for Infrared Astronomy (SOFIA). SOFIA was jointly operated by the Universities Space Research Association, Inc. (USRA), under NASA contract NNA17BF53C, and the Deutsches SOFIA Institut (DSI) under DLR contract 50 OK 2002 to the University of Stuttgart. Financial support for this work was provided by NASA through award 75\_0106 issued by USRA. Portions of this research were performed at the Jet Propulsion Laboratory, California Institute of Technology, under contract with the National Aeronautics and Space Administration and California Institute of Technology. S. L. N. is supported by NASA GO funding from the BAERI Cooperative Agreement \#80NSSC24M0065. J. P. F.  acknowledges funding support from Spanish Ministerio de Ciencia, Innovación, y Universidades through grants PID2023-147545NB-I00, PID2023-146056NB-C21, and PID2023-146056NB-C22.
% \end{acknowledgments}

\facilities{SOFIA(EXES)}

\software{\texttt{Astropy} \citep{AstropyCollaboration2013,AstropyCollaboration2018,AstropyCollaboration2022}, 
          \texttt{Astroquery} \citep{Ginsburg2019a},
           \texttt{HAPI} \citep{Kochanov2016},
           \texttt{Matplotlib} \citep{Hunter2007}, 
           \texttt{Numpy} \citep{Harris2020}, 
           \texttt{Scipy} \citep{Virtanen2020}. 
           }
\clearpage
\appendix
\restartappendixnumbering

\section{Observation Details}\label{ap:obs}

\begin{figure*}
\centering
\begin{tabular}{lr}
\includegraphics[scale=0.74]{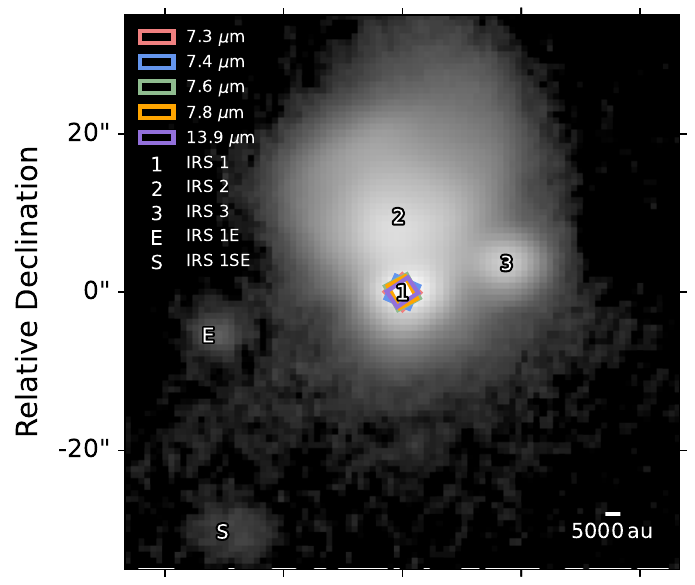}&
\includegraphics[scale=0.74]{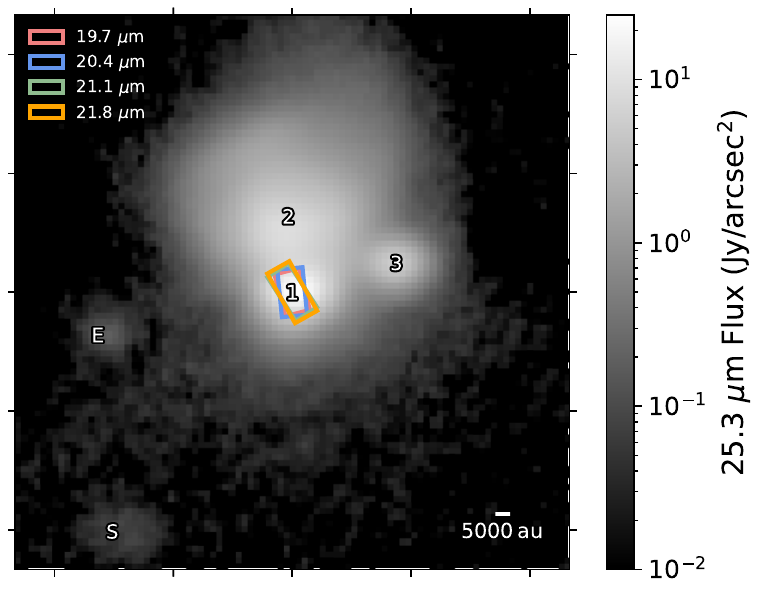}\\
\includegraphics[scale=0.74]{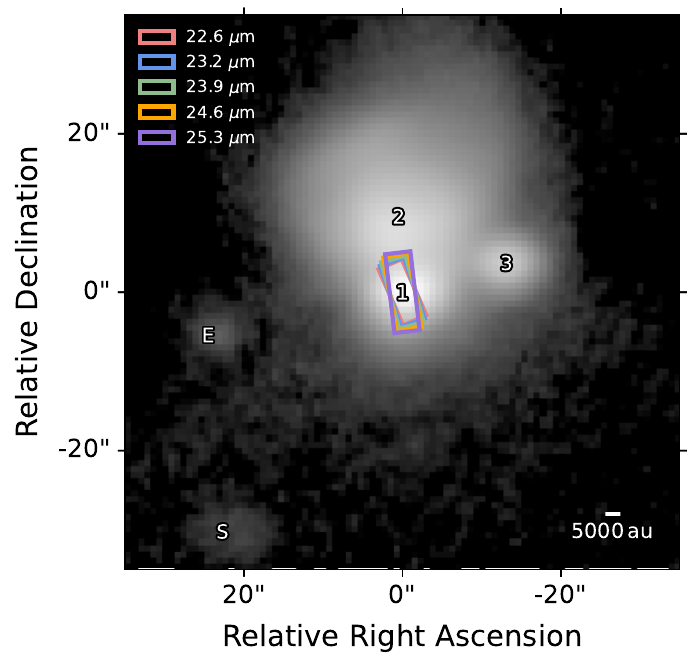}&
\includegraphics[scale=0.74]{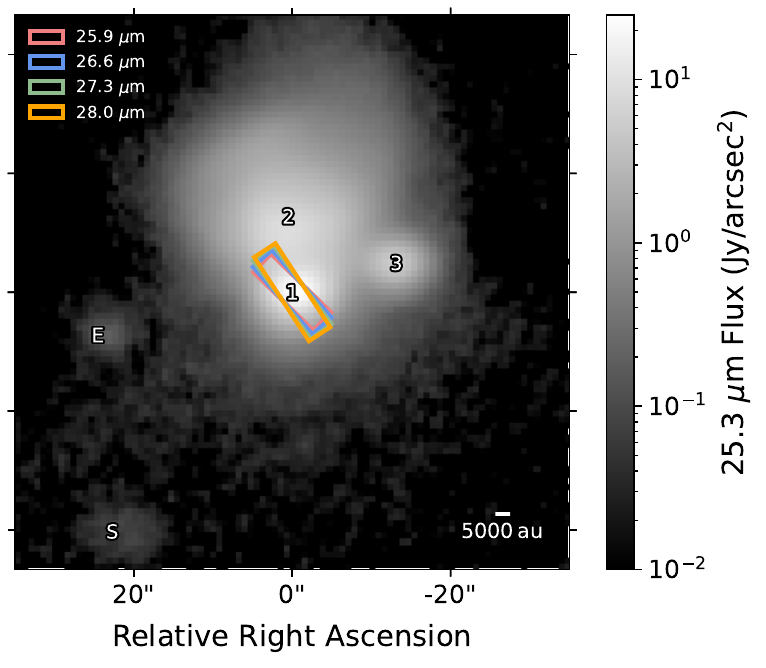}\\
\end{tabular}
\caption{The coloured boxes represent the beam configurations for EXES observations described in Table \ref{tab:obsexes}. The grey-scale map of the NGC 7538 IRS 1 region is SOFIA/FORCAST archival data at 25.3 \micron{}, as described in Figure \ref{fig:bigmap}. \label{fig:obsexes}}
\end{figure*}

In Table \ref{tab:obsexes} we list the SOFIA/EXES observations towards NGC 7538 IRS 1 and in Figure \ref{fig:obsexes} we show the slit position of each EXES observation overlaid on a SOFIA/FORCAST map of the NGC 7538 IRS 1 region.

%need startlong table to keep appendix single column
\startlongtable
\begin{deluxetable*}{cccccccc}
\tablecaption{Specifications for each setting taken with SOFIA/EXES (slit width: 3\farcs2) towards the massive protostar NGC 7538 IRS 1\label{tab:obsexes}}
\tablehead{\colhead{Setting} & \colhead{Species} & \colhead{Min $\lambda$} & \colhead{Max $\lambda$} &
\colhead{Date} & \colhead{Configuration} & \colhead{Slit Length} & \colhead{Exposure Time} \\
\colhead{(\micron)} & \colhead{} & \colhead{(\micron)} & \colhead{(\micron)} &
\colhead{(yyyy-mm-dd)} & \colhead{} & \colhead{(\arcsec)} & \colhead{(s)}}
\startdata
7.3 & \acet{} & 7.2 & 7.4 & 2018-10-30 & High-low & 3.2 & 4928 \\
7.4 & \acet{} & 7.3 & 7.5 & 2018-10-31 & High-low & 3.2 & 2496 \\
7.6 & \acet{} & 7.5 & 7.7 & 2018-10-31 & High-low & 3.4 & 2880 \\
7.8 & \acet{} & 7.7 & 7.9 & 2022-03-03 & High-low & 2.5 & 2560 \\
13.9 & \acet{} & 13.5 & 14.3 & 2022-05-03 & High-low & 2.4 & 640 \\
19.7 & \methanol{} & 19.4 & 20.1 & 2022-05-03 & High-low & 5.4 & 400 \\
20.4 & \methanol{}$^{\dagger}$ & 20.1 & 20.8 & 2022-05-03 & High-low & 6.0 & 400 \\
21.1 & \methanol{}$^{\dagger}$ & 20.8 & 21.5 & 2022-05-04 & High-low & 6.6 & 320 \\
21.8 & \methanol{} & 21.5 & 22.2 & 2022-05-04 & High-low & 7.2 & 320 \\
22.6 & \methanol{} & 22.2 & 22.9 & 2022-05-04 & High-low & 7.7 & 320 \\
23.2 & \methanol{}$^{\dagger}$ & 22.9 & 23.6 & 2022-05-04 & High-low & 8.1 & 320 \\
23.9 & \methanol{} & 23.5 & 24.2 & 2022-05-04 & High-low & 8.7 & 320 \\
24.6 & \methanol{} & 24.3 & 24.9 & 2022-05-04 & High-low & 9.4 & 320 \\
25.3 & \methanol{} & 24.9 & 25.6 & 2022-05-04 & High-low & 10.0 & 320 \\
25.9 & \methanol{} & 25.6 & 26.3 & 2022-05-05 & High-low & 10.6 & 320 \\
26.6 & \methanol{} & 26.2 & 26.9 & 2022-05-05 & High-low & 11.3 & 320 \\
27.3 & \methanol{} & 26.9 & 27.6 & 2022-05-05 & High-low & 12.3 & 320 \\
28.0 & \methanol{} & 27.6 & 28.3 & 2022-05-05 & High-low & 12.6 & 320 \\
\enddata
\tablecomments{Columns are, left to right: setting central wavelength, setting species, minimum wavelength, maximum wavelength, date of observation, EXES configuration, slit length, and exposure time. $^{\dagger}$Faint \methanol{} features are detected in these settings, but they are not strong enough against the noise level for analysis.}
\end{deluxetable*}

\section{25 \micron{} Methanol Line List}\label{ap:methll}

We have updated the \methanol{} line list reported in the laboratory work of \citet{Brauer2012} by adding the lower state energy values to all observed transitions. For this, we have adopted the lower state energy values from \citet{Moruzzi1995,Xu2008} and recent model calculations (private communication in 2022 with J. C. Pearson of the Jet Propulsion Laboratory) for the observed transitions by matching their quantum identification assignments (i.e. torsional quanta, J, K, and symmetry species) made by \citet{Brauer2012}. The updated line list covers the torsional fundamental ($\nu_{12}$--ground state), overtones (e.g. 2$\nu_{12}$--ground state, 3$\nu_{12}$--ground state), and hot bands (e.g. 2$\nu_{12}$--$\nu_{12}$, 3$\nu_{12}$--2$\nu_{12}$), as well as a few high J transitions from (3$\nu_{12}$--3$\nu_{12}$). When updated values were not available in the model predictions of J.C. Pearson, we have adopted the values from \citet{Xu2004}. It should be noted that these E'' values are calculated as distance from the Internal Rotation Barrier (127.97549 \ci{}) rather than from the absolute vibrational zero energy value, i.e. setting E'' = 0 for the A-species at J = K = 0 level. The non-zero K transitions are split into asymmetry doublets, labelled A+ and A$-$ \citep{Lees1968,Lees2020}.
%The updated line list covers the fundamental, overtones, and hot bands of the torsional mode, $\nu_{12}$, such as ($\nu_{12}$--ground state), (2$\nu_{12}$--ground state), (2$\nu_{12}$--$\nu_{12}$), (3$\nu_{12}$--2$\nu_{12}$) as well as high J transitions from (3$\nu_{12}$-–3$\nu_{12}$) band, etc. 

We have compiled an updated \methanol{} line list in the 300--500 \ci{} region, the beginning part of which is given in Table \ref{tab:methlist}. This line list will also be released with HITRAN2024. Note that we have included the spin factor of 4 from nuclear spin multiplicity in our calculations of the upper and lower statistical weights ($g_u$ and $g_l$ respectively) for consistency with its inclusion in the \citet{Villanueva2012} partition functions:
\begin{equation}
g_i=g_{i,angular}g_{spin}
\end{equation}
where $i=l$ or $u$ for lower and upper states, $g_{i,angular}=2J_i+1$ is the angular statistical weight, $J_i$ is the rotational total angular momentum, and $g_{spin}=4$ is the spin factor. Einstein $A$-coefficients are calculated according to Equation 20 in \citet{Simeckova2006}.

\startlongtable
\begin{deluxetable*}{rrrrrrrrrr}
\tablecaption{Line list for \methanol{} in the 300--500 \ci{} region, updated from \citet{Brauer2012} by adding the Einstein coefficient, lower and upper state energy, and the rotational statistical weights for the upper and lower states. We give the first five lines of the table here, and the full line list is available in the online version of this Letter. \label{tab:methlist}}
\tablehead{
\colhead{$\nu$}&\colhead{$S$}&\colhead{$A$}&\colhead{$E_l$}&\colhead{$V_u$}&\colhead{$V_l$}&\colhead{$Q_u$}&\colhead{$Q_l$}&\colhead{$g_u$}&\colhead{$g_l$}\\
\colhead{(cm$^{-1})$}&\colhead{(cm$^{-1}$/(molecule cm$^{-2}$))}&\colhead{(s$^{-1}$)}&\colhead{(cm$^{-1}$)}&\colhead{}&\colhead{}&\colhead{}&\colhead{}&\colhead{}&\colhead{}}
\startdata
300.031657&1.279E-21&0.123&325.0467&$\nu_{12}$&GS&20 1 E2&19 2 E2&164&156\\
300.162088&1.871E-21&0.190&250.053&$\nu_{12}$&GS&13 7 E1&12 6 E1&108&100\\
300.335032&1.028E-21&4.564&1091.9549&3$\nu_{12}$&3$\nu_{12}$&18 6 E1&17 5 E1&148&140\\
300.466076&3.504E-22&0.101&529.2193&2$\nu_{12}$&GS&18 7 A&19 8 A&148&156\\
300.479947&3.263E-21&0.605&402.1685&$\nu_{12}$&GS&15 9 E1&14 8 E1&124&116\\
\multicolumn{10}{c}{...}
\enddata
\tablecomments{$\nu$ is the rest wavenumber of the transition; $S$ is the transition intensity for a 100\% isotopologue sample (unlike HITRAN, not multiplied by terrestrial abundance); $A$ is the Einstein coefficient; $E_l$ is the lower-state energy; $V_u$ and $V_l$ are the upper and lower vibrational/torsional quanta; $Q_u$ and $Q_l$ are the rotational quanta consisting of J, K, and symmetry species (e.g. A, E); and $g_u$ and $g_l$ are the upper and lower state statistical weights. ``GS'' refers to ground state.}
\end{deluxetable*}

\section{Observed Transitions}\label{ap:lines}

In Table \ref{tab:abslines} we give the observed transitions and their inferred parameters for the measured \methanol{} and \acet{} lines observed towards NGC 7538 IRS 1 with SOFIA/EXES used in the rotation diagram analysis. In Table \ref{tab:blendlines} we give the properties of blended \methanol{} lines unsuitable for rotation diagram analysis.

Note that  we give the observed \vfwhm{} in Tables \ref{tab:rot} and \ref{tab:abslines}, and we may calculate the intrinsic \vfwhm{} by subtracting out the instrumental resolution $R$ via,
\begin{equation}
    \Delta v_{\mathrm{FWHM,intrinsic}}=\sqrt{\Delta v_{\mathrm{FWHM,observed}}^2-(c/R)^2},
\end{equation}
where $c$ is the speed of light.

\startlongtable
\begin{deluxetable*}{rrrrrrrrrl}
\tablecaption{Observed transitions and inferred parameters for molecular absorption lines observed with SOFIA/EXES towards NGC 7538 IRS 1.\label{tab:abslines}}
\tablehead{
\colhead{Transition} & \colhead{Wavelength} & \colhead{Wavenumber} & \colhead{$E_l/k_B$} & \colhead{$g_l$} & \colhead{$A$} & \colhead{\vlsr} & \colhead{\vfwhm} & \colhead{$\tau_0$} & \colhead{$N_l$} \\
\colhead{} & \colhead{(\micron)} & \colhead{(\ci)} & \colhead{(K)} & \colhead{} & \colhead{(\si)} & \colhead{(\kms)} & \colhead{(\kms)} & \colhead{} & \colhead{($\times10^{14}$\csi)}}
\startdata
\multicolumn{10}{c}{$\nu_{12}$ A-type \methanol{}}\\
R19R10&25.90879&385.969458&946.6&156&1.131&--58.1$\,\pm\,$0.9&6.3$\,\pm\,$3.1&0.009$\,\pm\,$0.003&0.08$\,\pm\,$0.05\\
R16R10&26.21962&381.393711&821.6&132&1.433&--58.0$\,\pm\,$0.4&6.6$\,\pm\,$1.3&0.026$\,\pm\,$0.004&0.17$\,\pm\,$0.04\\
R13R10&26.54084&376.7778&717.3&108&1.484&--58.9$\,\pm\,$0.3&6.8$\,\pm\,$1.0&0.037$\,\pm\,$0.004&0.22$\,\pm\,$0.04\\
R10R10&26.87297&372.121122&633.9&84&1.612&--59.2$\,\pm\,$0.3&7.5$\,\pm\,$0.8&0.040$\,\pm\,$0.003&0.24$\,\pm\,$0.03\\
\hline
\multicolumn{10}{c}{$2\nu_{12}$ A-type \methanol{}}\\
R15R0&20.02856&499.286925&278.2&124&0.114&--55.8$\,\pm\,$1.1&8.3$\,\pm\,$3.5&0.028$\,\pm\,$0.007&6.41$\,\pm\,$3.09\\
R13R3&24.13029&414.416923&261.0&108&0.103&--58.8$\,\pm\,$0.4&6.0$\,\pm\,$1.5&0.034$\,\pm\,$0.006&3.54$\,\pm\,$1.14\\
R11R3&24.30928&411.365588&203.0&92&0.193&--58.5$\,\pm\,$0.3&5.1$\,\pm\,$0.7&0.049$\,\pm\,$0.005&2.22$\,\pm\,$0.34\\
$^{+-}$R8R3&24.58616&406.73297&133.4&68&0.107&--58.0$\,\pm\,$0.3&7.2$\,\pm\,$1.1&0.043$\,\pm\,$0.005&4.67$\,\pm\,$0.85\\
R7R3&24.68066&405.175538&114.8&60&0.222&--58.0$\,\pm\,$0.3&6.3$\,\pm\,$0.9&0.060$\,\pm\,$0.006&2.70$\,\pm\,$0.50\\
R5R3&24.87310&402.040837&84.6&44&0.256&--58.2$\,\pm\,$0.2&5.8$\,\pm\,$0.8&0.065$\,\pm\,$0.006&2.14$\,\pm\,$0.37\\
Q6P4&24.91088&401.431007&129.1&52&0.120&--57.3$\,\pm\,$0.5&5.3$\,\pm\,$1.3&0.030$\,\pm\,$0.006&2.32$\,\pm\,$0.59\\
Q7P4&24.91349&401.388966&145.3&60&0.131&--57.9$\,\pm\,$0.4&4.8$\,\pm\,$1.0&0.039$\,\pm\,$0.006&2.46$\,\pm\,$0.52\\
Q8P4&24.91650&401.340525&163.9&68&0.142&--57.7$\,\pm\,$0.6&6.8$\,\pm\,$2.0&0.026$\,\pm\,$0.005&2.15$\,\pm\,$0.83\\
Q9P4&24.91990&401.285745&184.8&76&0.147&--57.5$\,\pm\,$0.7&8.0$\,\pm\,$2.4&0.027$\,\pm\,$0.006&2.58$\,\pm\,$0.97\\
Q10P4&24.92372&401.224221&208.0&84&0.146&--59.0$\,\pm\,$0.8&6.6$\,\pm\,$2.0&0.039$\,\pm\,$0.006&3.06$\,\pm\,$0.96\\
Q11P4&24.92796&401.155975&233.5&92&0.146&--59.7$\,\pm\,$0.7&11.5$\,\pm\,$4.7&0.036$\,\pm\,$0.013&4.93$\,\pm\,$3.62\\
Q12P4&24.93263&401.080795&261.4&100&0.167&--55.7$\,\pm\,$0.7&9.5$\,\pm\,$3.4&0.027$\,\pm\,$0.007&2.62$\,\pm\,$1.41\\
Q16P4&24.95596&400.705879&395.9&132&0.152&--57.3$\,\pm\,$0.5&6.6$\,\pm\,$2.3&0.030$\,\pm\,$0.008&2.22$\,\pm\,$1.22\\
R4R3&24.97101&400.464416&73.0&36&0.314&--57.4$\,\pm\,$0.2&4.9$\,\pm\,$0.6&0.070$\,\pm\,$0.007&1.53$\,\pm\,$0.26\\
P4P4&25.31212&395.067622&103.6&36&0.370&--56.9$\,\pm\,$0.5&5.7$\,\pm\,$1.3&0.049$\,\pm\,$0.009&1.60$\,\pm\,$0.41\\
P5P4&25.41743&393.430859&115.2&44&0.247&--58.1$\,\pm\,$0.2&5.8$\,\pm\,$0.8&0.059$\,\pm\,$0.006&2.72$\,\pm\,$0.50\\
$^{+-}$Q9R3&25.49617&392.215737&154.2&76&0.089&--59.7$\,\pm\,$0.3&7.5$\,\pm\,$1.2&0.053$\,\pm\,$0.006&7.23$\,\pm\,$1.65\\
$^{+-}$Q11R3&25.50462&392.08588&203.0&92&0.095&--58.5$\,\pm\,$0.2&6.2$\,\pm\,$0.4&0.058$\,\pm\,$0.003&6.07$\,\pm\,$0.63\\
Q13R3&25.51446&391.934584&261.0&108&0.090&--56.0$\,\pm\,$0.3&6.9$\,\pm\,$0.9&0.053$\,\pm\,$0.006&6.64$\,\pm\,$0.93\\
P7P4&25.63184&390.13983&145.3&60&0.176&--57.9$\,\pm\,$0.3&5.3$\,\pm\,$0.7&0.038$\,\pm\,$0.004&2.09$\,\pm\,$0.39\\
P8P4&25.74102&388.48505&163.9&68&0.162&--57.1$\,\pm\,$0.4&6.9$\,\pm\,$1.4&0.034$\,\pm\,$0.005&2.58$\,\pm\,$0.63\\
P10P4&25.96343&385.157121&208.0&84&0.144&--57.7$\,\pm\,$0.3&4.0$\,\pm\,$0.7&0.059$\,\pm\,$0.009&2.80$\,\pm\,$0.56\\
P11P4&26.07676&383.483167&233.5&92&0.141&--58.2$\,\pm\,$0.4&8.7$\,\pm\,$1.6&0.027$\,\pm\,$0.003&2.72$\,\pm\,$0.64\\
P15P4&26.54486&376.720746&358.8&124&0.109&--58.4$\,\pm\,$0.6&4.6$\,\pm\,$1.7&0.018$\,\pm\,$0.005&1.17$\,\pm\,$0.51\\
\hline
\multicolumn{10}{c}{$\nu_{12}$ E-type \methanol{}}\\
R16R9&26.49462&377.435182&724.1&132&0.429&--60.5$\,\pm\,$0.7&6.7$\,\pm\,$2.4&0.013$\,\pm\,$0.003&0.27$\,\pm\,$0.12\\
R11R11&27.66277&361.496685&766.5&92&0.962&--57.0$\,\pm\,$0.8&6.1$\,\pm\,$2.1&0.026$\,\pm\,$0.005&0.19$\,\pm\,$0.07\\
\hline
\multicolumn{10}{c}{$2\nu_{12}$ E-type \methanol{}}\\
R7R6&20.03195&499.202436&243.8&60&0.204&--57.5$\,\pm\,$0.6&5.6$\,\pm\,$1.6&0.042$\,\pm\,$0.010&3.36$\,\pm\,$1.01\\
R6R1&21.95320&455.514516&54.3&52&0.078&--58.3$\,\pm\,$0.6&4.4$\,\pm\,$1.6&0.044$\,\pm\,$0.013&5.41$\,\pm\,$2.07\\
R1R1&22.34181&447.591215&7.9&12&0.119&--57.0$\,\pm\,$0.4&4.6$\,\pm\,$1.1&0.035$\,\pm\,$0.006&1.94$\,\pm\,$0.49\\
R8R8&22.34762&447.475008&405.0&68&0.266&--55.4$\,\pm\,$0.8&6.9$\,\pm\,$2.0&0.022$\,\pm\,$0.005&1.21$\,\pm\,$0.34\\
Q4R1&22.50611&444.323822&28.8&36&0.112&--57.6$\,\pm\,$0.4&6.2$\,\pm\,$1.2&0.042$\,\pm\,$0.006&5.45$\,\pm\,$1.13\\
Q9R1&22.51489&444.150507&110.0&76&0.094&--59.5$\,\pm\,$0.6&8.0$\,\pm\,$1.8&0.034$\,\pm\,$0.006&6.76$\,\pm\,$1.79\\
Q11R1&22.51904&444.068745&158.6&92&0.099&--57.5$\,\pm\,$0.4&5.0$\,\pm\,$1.1&0.044$\,\pm\,$0.008&5.27$\,\pm\,$1.76\\
R13R2&22.81625&438.284146&233.6&108&0.087&--56.6$\,\pm\,$0.7&7.2$\,\pm\,$2.1&0.036$\,\pm\,$0.008&6.19$\,\pm\,$2.20\\
R4R2&23.56054&424.438567&45.5&36&0.121&--58.5$\,\pm\,$0.5&7.9$\,\pm\,$1.9&0.035$\,\pm\,$0.006&3.77$\,\pm\,$1.14\\
R2R2&23.73791&421.267113&29.2&20&0.167&--58.1$\,\pm\,$0.2&4.5$\,\pm\,$0.6&0.065$\,\pm\,$0.006&2.48$\,\pm\,$0.38\\
Q4R2&24.01528&416.401535&45.5&36&0.073&--57.4$\,\pm\,$0.5&7.0$\,\pm\,$1.5&0.037$\,\pm\,$0.006&6.83$\,\pm\,$1.61\\
Q5R2&24.01712&416.369609&57.1&44&0.094&--59.6$\,\pm\,$0.4&4.5$\,\pm\,$1.1&0.047$\,\pm\,$0.009&4.30$\,\pm\,$1.07\\
Q6R2&24.01940&416.330141&71.0&52&0.105&--57.6$\,\pm\,$0.5&5.7$\,\pm\,$1.2&0.038$\,\pm\,$0.007&3.93$\,\pm\,$0.82\\
Q11R2&24.03854&415.998666&175.5&92&0.102&--58.7$\,\pm\,$0.6&5.7$\,\pm\,$1.6&0.023$\,\pm\,$0.005&2.47$\,\pm\,$0.73\\
R14R4&24.84242&402.537276&331.5&116&0.050&--57.5$\,\pm\,$0.7&8.3$\,\pm\,$2.0&0.023$\,\pm\,$0.004&6.34$\,\pm\,$1.60\\
R16P3&24.95712&400.687314&363.4&132&0.039&--56.1$\,\pm\,$0.4&5.8$\,\pm\,$1.2&0.020$\,\pm\,$0.003&4.75$\,\pm\,$1.09\\
R8R4&25.42375&393.332938&171.5&68&0.056&--57.6$\,\pm\,$0.3&6.0$\,\pm\,$0.8&0.037$\,\pm\,$0.004&5.73$\,\pm\,$0.88\\
R5R4&25.73036&388.645902&122.7&44&0.089&--56.7$\,\pm\,$0.2&7.3$\,\pm\,$1.1&0.043$\,\pm\,$0.005&4.75$\,\pm\,$1.05\\
Q12R4&26.41208&378.614622&268.9&100&0.046&--59.0$\,\pm\,$0.5&5.5$\,\pm\,$1.4&0.025$\,\pm\,$0.005&4.34$\,\pm\,$1.19\\
Q5P3&26.72335&374.20456&82.5&44&0.068&--57.5$\,\pm\,$0.4&5.5$\,\pm\,$1.1&0.029$\,\pm\,$0.005&3.34$\,\pm\,$1.03\\
Q6P3&26.72595&374.168206&96.5&52&0.073&--59.3$\,\pm\,$0.4&6.7$\,\pm\,$1.0&0.032$\,\pm\,$0.004&4.04$\,\pm\,$0.65\\
Q7P3&26.72901&374.125404&112.7&60&0.081&--58.3$\,\pm\,$0.3&5.7$\,\pm\,$1.3&0.040$\,\pm\,$0.006&3.96$\,\pm\,$1.12\\
Q9P3&26.73657&374.019567&152.2&76&0.092&--57.5$\,\pm\,$0.3&5.3$\,\pm\,$0.7&0.045$\,\pm\,$0.005&3.64$\,\pm\,$0.47\\
Q10P3&26.74119&373.954867&175.4&84&0.077&--59.0$\,\pm\,$0.4&5.3$\,\pm\,$1.5&0.039$\,\pm\,$0.008&3.80$\,\pm\,$1.40\\
Q14P3&26.76556&373.614454&291.5&116&0.085&--59.5$\,\pm\,$0.3&5.9$\,\pm\,$0.9&0.033$\,\pm\,$0.004&3.16$\,\pm\,$0.62\\
R13P2&26.76755&373.586738&237.3&108&0.037&--58.3$\,\pm\,$0.3&4.2$\,\pm\,$0.8&0.027$\,\pm\,$0.004&3.96$\,\pm\,$0.89\\
Q15P3&26.77330&373.506377&326.3&124&0.087&--58.8$\,\pm\,$0.5&6.0$\,\pm\,$1.4&0.022$\,\pm\,$0.004&2.09$\,\pm\,$0.57\\
Q16P3&26.78184&373.387285&363.4&132&0.079&--58.4$\,\pm\,$0.7&8.0$\,\pm\,$1.9&0.024$\,\pm\,$0.004&3.43$\,\pm\,$0.89\\
P3P3&27.06831&369.435742&61.6&28&0.188&--58.3$\,\pm\,$0.3&8.7$\,\pm\,$1.0&0.047$\,\pm\,$0.004&4.11$\,\pm\,$0.57\\
R9P2&27.20720&367.54981&130.4&76&0.029&--59.3$\,\pm\,$0.3&6.2$\,\pm\,$0.9&0.033$\,\pm\,$0.004&8.67$\,\pm\,$1.60\\
\hline
\multicolumn{10}{c}{$\nu_{5}$ o-\acet{}}\\
R1e&13.62659&733.85945&3.4&9&3.693&--59.0$\,\pm\,$0.2&7.6$\,\pm\,$0.5&0.132$\,\pm\,$0.007&1.72$\,\pm\,$0.13\\
Q21e&13.67576&731.22084&781.6&129&6.111&--57.9$\,\pm\,$0.2&8.5$\,\pm\,$0.7&0.079$\,\pm\,$0.005&1.15$\,\pm\,$0.14\\
Q19e&13.68260&730.85539&643.0&117&6.102&--57.6$\,\pm\,$0.2&4.8$\,\pm\,$0.4&0.103$\,\pm\,$0.007&0.84$\,\pm\,$0.09\\
Q17e&13.68878&730.52509&517.8&105&6.094&--57.9$\,\pm\,$0.2&6.7$\,\pm\,$0.4&0.138$\,\pm\,$0.007&1.59$\,\pm\,$0.11\\
Q15e&13.69431&730.23009&406.2&93&6.086&--58.3$\,\pm\,$0.2&6.5$\,\pm\,$0.5&0.127$\,\pm\,$0.007&1.41$\,\pm\,$0.12\\
Q7e&13.70979&729.40565&94.8&45&6.067&--59.2$\,\pm\,$0.1&6.4$\,\pm\,$0.3&0.181$\,\pm\,$0.006&1.97$\,\pm\,$0.10\\
Q3e&13.71351&729.20806&20.3&21&6.061&--58.3$\,\pm\,$0.1&6.1$\,\pm\,$0.5&0.185$\,\pm\,$0.011&1.92$\,\pm\,$0.26\\
P7e&14.03165&712.67472&94.8&45&2.649&--59.4$\,\pm\,$0.2&4.7$\,\pm\,$0.6&0.091$\,\pm\,$0.009&1.78$\,\pm\,$0.25\\
\hline
\multicolumn{10}{c}{$\nu_{5}$ p-\acet{}}\\
R4e&13.49685&740.91365&33.9&9&3.447&--56.4$\,\pm\,$0.2&7.6$\,\pm\,$1.1&0.120$\,\pm\,$0.013&2.37$\,\pm\,$0.53\\
R2e&13.58306&736.21139&10.2&5&3.548&--58.8$\,\pm\,$0.2&7.1$\,\pm\,$1.0&0.112$\,\pm\,$0.011&1.72$\,\pm\,$0.35\\
Q18e&13.68577&730.68584&578.7&37&6.099&--58.1$\,\pm\,$0.3&8.0$\,\pm\,$0.7&0.079$\,\pm\,$0.005&1.07$\,\pm\,$0.10\\
Q6e&13.71097&729.3428&71.1&13&6.065&--58.3$\,\pm\,$0.1&7.3$\,\pm\,$1.1&0.133$\,\pm\,$0.018&1.66$\,\pm\,$0.47\\
P2e&13.80363&724.4472&10.2&5&1.986&--58.1$\,\pm\,$0.2&5.9$\,\pm\,$0.6&0.061$\,\pm\,$0.005&3.08$\,\pm\,$0.35\\
\hline
\multicolumn{10}{c}{$\nu_{4}+\nu_{5}$ o-\acet{}}\\
R21e&7.23976&1381.26196&781.6&129&2.218&--58.3$\,\pm\,$0.4&4.4$\,\pm\,$1.1&0.038$\,\pm\,$0.008&5.03$\,\pm\,$1.28\\
R19e&7.26542&1376.381985&643.0&117&2.218&--59.4$\,\pm\,$0.8&8.8$\,\pm\,$2.3&0.025$\,\pm\,$0.005&6.49$\,\pm\,$1.67\\
R17e&7.29131&1371.495566&517.8&105&2.213&--57.5$\,\pm\,$0.2&6.8$\,\pm\,$0.5&0.070$\,\pm\,$0.004&14.02$\,\pm\,$1.38\\
R15e&7.31738&1366.608598&406.2&93&2.202&--57.1$\,\pm\,$0.2&9.5$\,\pm\,$0.9&0.079$\,\pm\,$0.005&21.86$\,\pm\,$2.91\\
R13e&7.34362&1361.726778&308.0&81&2.186&--54.5$\,\pm\,$0.3&10.6$\,\pm\,$1.0&0.079$\,\pm\,$0.005&24.08$\,\pm\,$2.80\\
R11e&7.36998&1356.855432&223.4&69&2.163&--57.2$\,\pm\,$0.3&11.1$\,\pm\,$1.9&0.067$\,\pm\,$0.009&21.23$\,\pm\,$6.21\\
R7e&7.42301&1347.162859&94.8&45&2.097&--58.5$\,\pm\,$0.5&11.9$\,\pm\,$2.3&0.063$\,\pm\,$0.009&20.68$\,\pm\,$5.98\\
P3e&7.56981&1321.036604&20.3&21&2.578&--57.1$\,\pm\,$0.3&9.3$\,\pm\,$1.0&0.054$\,\pm\,$0.004&16.80$\,\pm\,$2.32\\
P5e&7.59657&1316.383626&50.8&33&2.371&--58.3$\,\pm\,$0.4&11.0$\,\pm\,$1.3&0.076$\,\pm\,$0.007&26.18$\,\pm\,$4.12\\
P9e&7.65014&1307.16496&152.3&57&2.228&--57.2$\,\pm\,$0.4&8.7$\,\pm\,$1.1&0.067$\,\pm\,$0.007&17.27$\,\pm\,$2.46\\
P15e&7.73084&1293.521104&406.2&93&2.124&--57.6$\,\pm\,$0.3&8.3$\,\pm\,$1.0&0.073$\,\pm\,$0.006&17.68$\,\pm\,$2.56\\
P17e&7.75790&1289.008551&517.8&105&2.094&--56.8$\,\pm\,$0.4&9.1$\,\pm\,$0.9&0.067$\,\pm\,$0.004&17.73$\,\pm\,$2.06\\
\hline
\multicolumn{10}{c}{$\nu_{4}+\nu_{5}$ p-\acet{}}\\
R20e&7.25256&1378.823149&710.6&41&2.220&--55.6$\,\pm\,$0.5&6.6$\,\pm\,$1.5&0.026$\,\pm\,$0.004&5.24$\,\pm\,$1.42\\
R16e&7.30433&1369.051788&460.3&33&2.208&--58.0$\,\pm\,$0.4&6.6$\,\pm\,$1.2&0.049$\,\pm\,$0.006&9.59$\,\pm\,$1.96\\
R14e&7.33048&1364.166701&355.4&29&2.194&--58.4$\,\pm\,$0.3&7.9$\,\pm\,$1.2&0.045$\,\pm\,$0.005&10.21$\,\pm\,$2.01\\
R12e&7.35678&1359.289481&264.0&25&2.175&--57.3$\,\pm\,$0.3&7.4$\,\pm\,$1.3&0.044$\,\pm\,$0.005&9.28$\,\pm\,$2.21\\
R10e&7.38321&1354.425216&186.2&21&2.150&--57.3$\,\pm\,$0.4&7.1$\,\pm\,$1.4&0.036$\,\pm\,$0.005&7.34$\,\pm\,$1.76\\
R8e&7.40972&1349.578441&121.9&17&2.118&--55.6$\,\pm\,$0.3&6.3$\,\pm\,$0.8&0.056$\,\pm\,$0.006&9.78$\,\pm\,$1.81\\
R6e&7.43631&1344.753064&71.1&13&2.073&--57.5$\,\pm\,$0.4&8.8$\,\pm\,$1.2&0.061$\,\pm\,$0.006&14.65$\,\pm\,$2.28\\
\enddata
\tablecomments{Wavelength and wavenumber are the rest value for each transition, $E_l$ is the energy level of the lower state, $k_B$ is the Boltzmann constant, $g_l$ is the lower statistical weight, $A$ is the Einstein coefficient, \vlsr\ is the observed local standard of rest velocity at the lines's centre, \vfwhm\ is the observed full-width half-maximum, $\tau_0$ is the observed optical depth, and $N_l$ is the estimated column density of the lines. Data in the first six columns for \methanol{} are from the line list described in Appendix \ref{ap:methll} and for \acet{} the HITRAN database \citep{Gordon2022}. Transition labels for \methanol{} follow the format: $\Delta$J$\Delta$K, where J and K are the lower state rotational quanta and P, Q, R are for $\Delta$J or $\Delta$K = $-1$, 0, 1. $^{+-}$Transitions split into A- and A+ initial states, which we treat as a single lines.}
\end{deluxetable*}

%todo check in final paper type and band
\startlongtable
\begin{deluxetable*}{rrrrrrrrrrrr}
\tablecaption{SOFIA/EXES observed lines towards NGC 7538 IRS 1 for strong \methanol{} absorption lines not fit due to blending. \label{tab:blendlines}}
\tablehead{
\colhead{Transition} & \colhead{Wavelength} & \colhead{Wavenumber} & \colhead{$E_l/k_B$} & \colhead{$g_l$} & \colhead{$A$} & \colhead{Type} &\colhead{Band}\\
\colhead{} & \colhead{(\micron)} & \colhead{(\ci)} & \colhead{(K)} & \colhead{} & \colhead{(\si)}& \colhead{} & \colhead{} }
\startdata
R7P1&28.07464&356.193305&79.0&60&0.020&A&$2\nu_{12}$\\
P11P3&28.07352&356.207534&200.9&92&0.056&E&$2\nu_{12}$\\[1.5mm]
Q10R4&26.40230&378.754876&215.5&84&0.012&E&$2\nu_{12}$\\
R12P9&26.40230&378.754876&583.7&100&0.033&E&$2\nu_{12}$\\[1.5mm]
Q24R3&25.59842&390.649102&745.5&196&0.080&A&$2\nu_{12}$\\
Q23R3&25.59796&390.656177&690.1&188&0.107&A&$2\nu_{12}$\\[1.5mm]
R7R4&25.52473&391.776877&152.9&60&0.056&E&$2\nu_{12}$\\
P6P4&25.52399&391.788283&129.1&52&0.208&A&$2\nu_{12}$\\[1.5mm]
Q12R3&25.50954&392.010231&230.8&100&0.141&A&$2\nu_{12}$\\
R23R10&25.50954&392.010231&1145.6&188&1.586&A&$\nu_{12}$\\[1.5mm]
$^{+-}$Q10R3&25.50016&392.154437&177.5&84&0.063&A&$2\nu_{12}$\\
Q14R1&25.50016&392.154437&264.8&116&0.062&E&$2\nu_{12}$\\[1.5mm]
Q13R2&24.05031&415.795013&233.6&108&0.109&E&$2\nu_{12}$\\
R8P4&24.04986&415.802906&163.9&68&0.031&A&$2\nu_{12}$\\[1.5mm]
Q13R9&24.04413&415.901963&619.9&108&0.211&E&$2\nu_{12}$\\
Q12R2&24.04413&415.901963&203.4&100&0.056&E&$2\nu_{12}$\\[1.5mm]
$^{\mathrm{H}_2\mathrm{O}}$R21R3&23.45842&426.286194&585.7&172&0.115&A&$2\nu_{12}$\\[1.5mm]
Q24R1&22.51360&444.175922&698.9&196&0.117&E&$2\nu_{12}$\\
R17R2&22.51337&444.180517&377.6&140&0.067&E&$2\nu_{12}$\\
Q8R1&22.51285&444.190789&89.1&68&0.106&E&$2\nu_{12}$\\[1.5mm]
\enddata
\tablecomments{Columns are explained in Table \ref{tab:abslines}, with additional columns for type and band. Spaces separate blended groups. $^{+-}$: initial state split between A+ and A-; $^{\mathrm{H}_2\mathrm{O}}$: Blended with an \water{} line.}
\end{deluxetable*}

\section{Curve of Growth Analysis}\label{ap:cog}

\begin{figure*}
\centering
\gridline{\fig{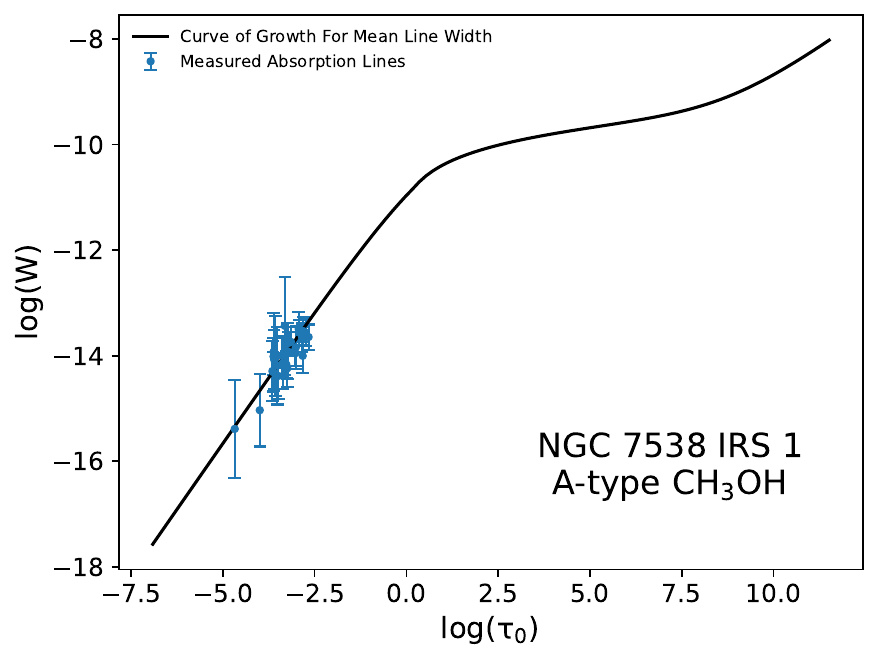}{0.49\textwidth}{}\hspace{-5mm}
          \fig{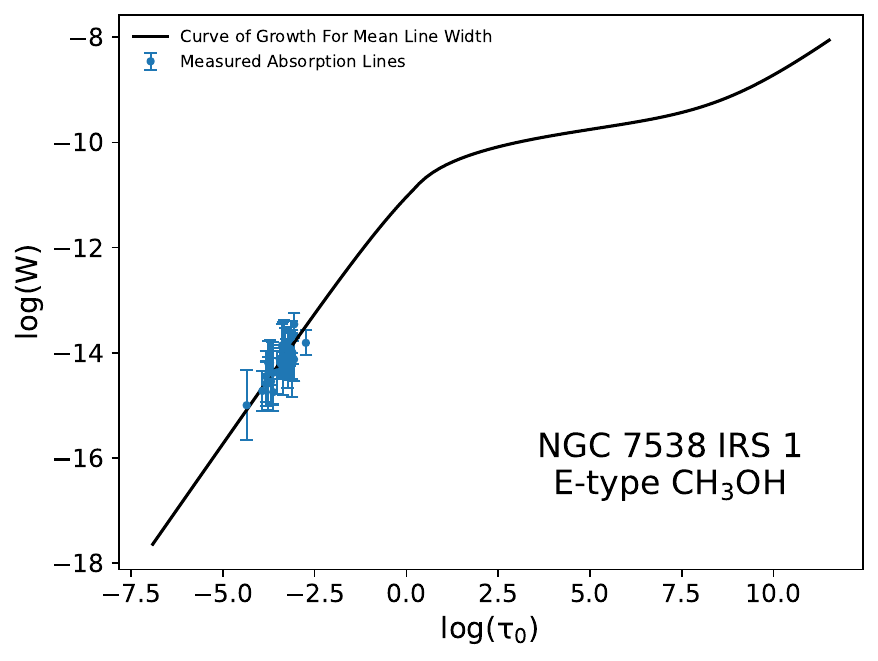}{0.49\textwidth}{}}
\vspace{-10mm}
\gridline{\fig{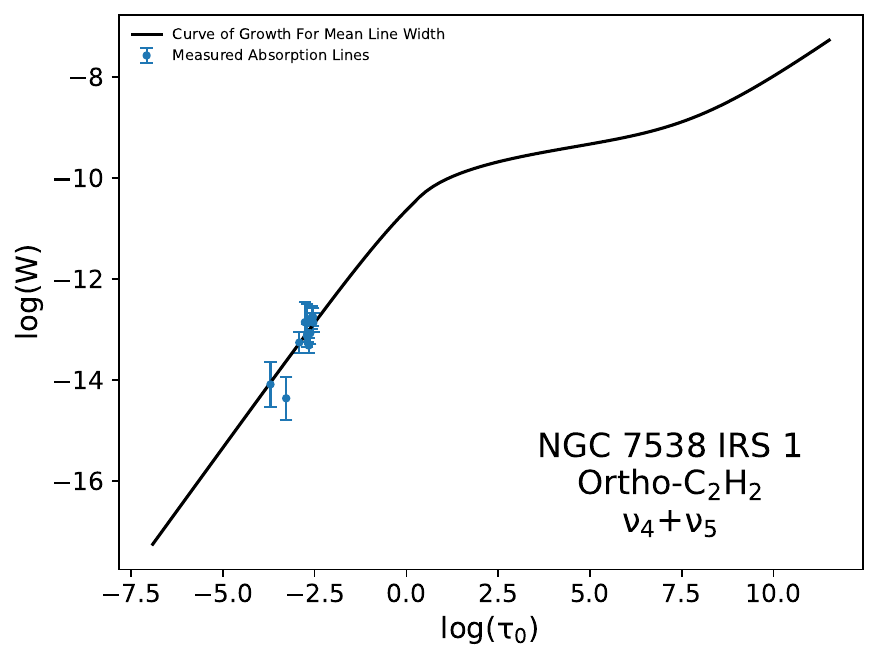}{0.49\textwidth}{}\hspace{-5mm}
          \fig{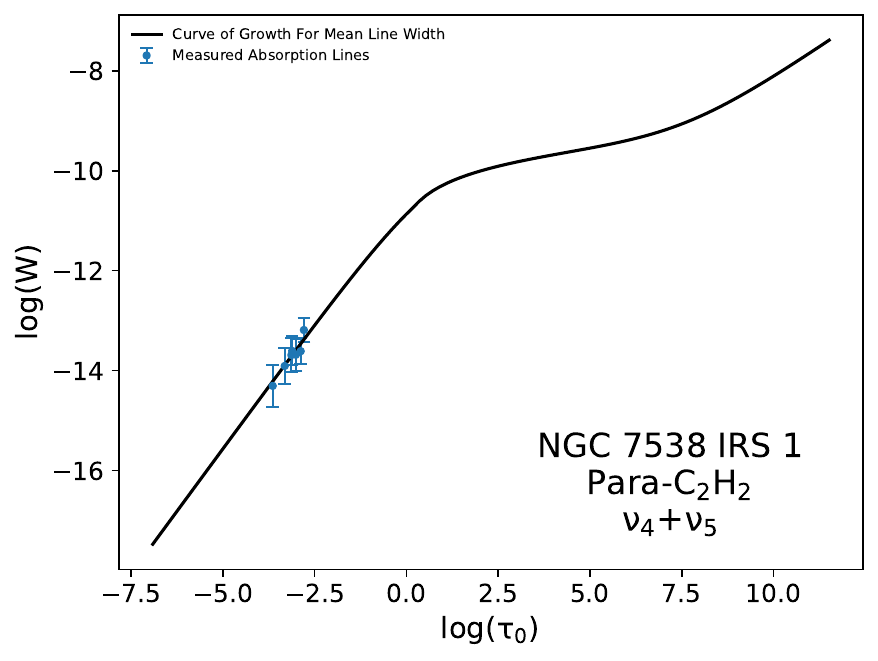}{0.49\textwidth}{}}
          \vspace{-10mm}
\gridline{\fig{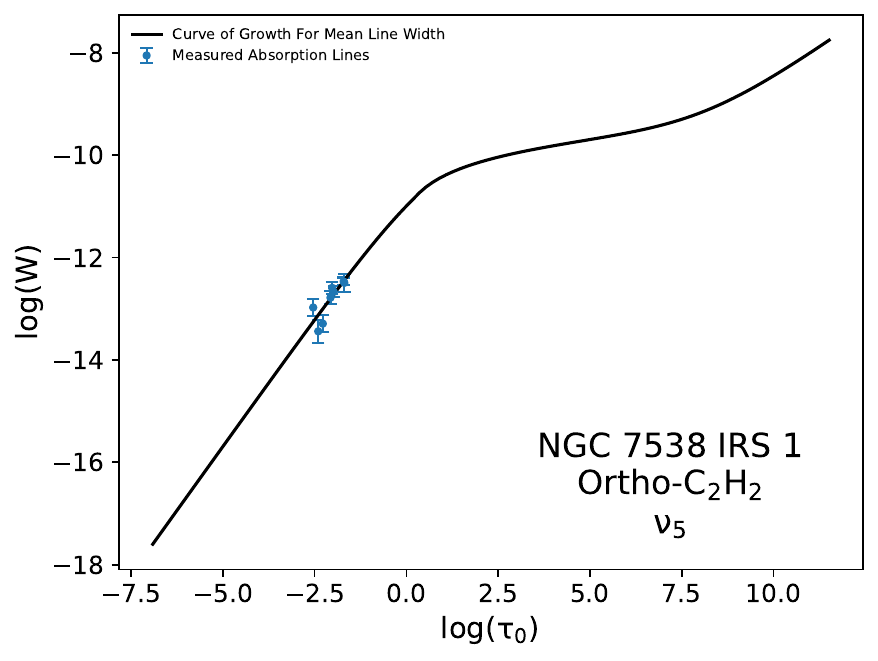}{0.49\textwidth}{}\hspace{-5mm}
          \fig{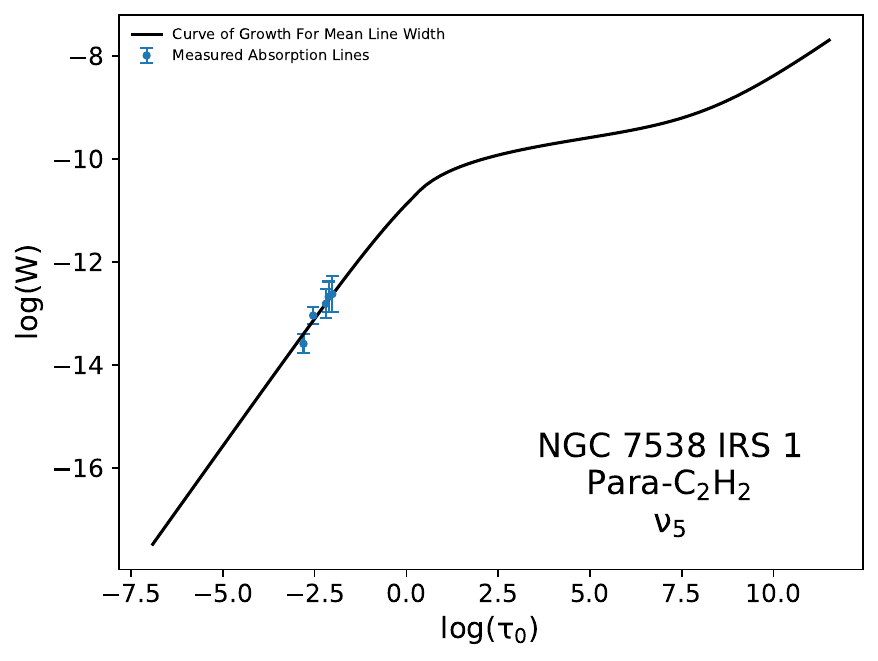}{0.49\textwidth}{}}
\caption{Curve of growth plots for absorption lines measured by SOFIA/EXES towards NGC 7538 IRS 1 for separate \methanol{} and \acet{} groups. \label{fig:curvegrow}}
\end{figure*}

We carried out a curve of growth analysis on EXES measurements of \methanol{} and \acet{} towards NGC 7538 IRS 1 to determine if the absorption lines are in the optically thin or saturated regime. The procedure is described in Chapter 9 of \citet{Draine2011}. Figure \ref{fig:curvegrow} gives the optical depth at line centre, $\tau_0$ (Equation 9.9 in \citealt{Draine2011}, neglecting the stimulated emission term), versus the dimensionless equivalent width, $W$, which is dependent on $\tau_0$ and the line width \vfwhm{} (Equation 9.27 in \citealt{Draine2011}). In order to determine in which regime the lines belongs, we plotted $W$ with the mean \vfwhm{}. All molecular species fall in the flat portion of the curve of growth, where the line width increases with increasing depth. Therefore, our lines are optically thin.

\bibliography{zotero}{}

\begin{thebibliography}{}
\expandafter\ifx\csname natexlab\endcsname\relax\def\natexlab#1{#1}\fi
\providecommand{\url}[1]{\href{#1}{#1}}
\providecommand{\dodoi}[1]{doi:~\href{http://doi.org/#1}{\nolinkurl{#1}}}
\providecommand{\doeprint}[1]{\href{http://ascl.net/#1}{\nolinkurl{http://ascl.net/#1}}}
\providecommand{\doarXiv}[1]{\href{https://arxiv.org/abs/#1}{\nolinkurl{https://arxiv.org/abs/#1}}}

\bibitem[{Aikawa {et~al.}(2020)Aikawa, Furuya, Yamamoto, \& Sakai}]{Aikawa2020}
Aikawa, Y., Furuya, K., Yamamoto, S., \& Sakai, N. 2020, The Astrophysical Journal, 897, 110, \dodoi{10.3847/1538-4357/ab994a}

\bibitem[{An {et~al.}(2017)An, Sellgren, Boogert, Ramírez, \& Pyo}]{An2017}
An, D., Sellgren, K., Boogert, A. C.~A., Ramírez, S.~V., \& Pyo, T.-S. 2017, The Astrophysical Journal, 843, L36, \dodoi{10.3847/2041-8213/aa7cfe}

\bibitem[{Armitage(2024)}]{Armitage2024}
Armitage, P.~J. 2024, Planet formation theory: an overview, \dodoi{10.48550/arXiv.2412.11064}

\bibitem[{{Astropy Collaboration} {et~al.}(2013){Astropy Collaboration}, Robitaille, Tollerud, Greenfield, Droettboom, Bray, Aldcroft, Davis, Ginsburg, Price-Whelan, Kerzendorf, Conley, Crighton, Barbary, Muna, Ferguson, Grollier, Parikh, Nair, Unther, Deil, Woillez, Conseil, Kramer, Turner, Singer, Fox, Weaver, Zabalza, Edwards, Azalee~Bostroem, Burke, Casey, Crawford, Dencheva, Ely, Jenness, Labrie, Lim, Pierfederici, Pontzen, Ptak, Refsdal, Servillat, \& Streicher}]{AstropyCollaboration2013}
{Astropy Collaboration}, Robitaille, T.~P., Tollerud, E.~J., {et~al.} 2013, Astronomy and Astrophysics, 558, A33, \dodoi{10.1051/0004-6361/201322068}

\bibitem[{{Astropy Collaboration} {et~al.}(2018){Astropy Collaboration}, Price-Whelan, Sipőcz, Günther, Lim, Crawford, Conseil, Shupe, Craig, Dencheva, Ginsburg, VanderPlas, Bradley, Pérez-Suárez, de~Val-Borro, Aldcroft, Cruz, Robitaille, Tollerud, Ardelean, Babej, Bach, Bachetti, Bakanov, Bamford, Barentsen, Barmby, Baumbach, Berry, Biscani, Boquien, Bostroem, Bouma, Brammer, Bray, Breytenbach, Buddelmeijer, Burke, Calderone, Cano~Rodríguez, Cara, Cardoso, Cheedella, Copin, Corrales, Crichton, D'Avella, Deil, Depagne, Dietrich, Donath, Droettboom, Earl, Erben, Fabbro, Ferreira, Finethy, Fox, Garrison, Gibbons, Goldstein, Gommers, Greco, Greenfield, Groener, Grollier, Hagen, Hirst, Homeier, Horton, Hosseinzadeh, Hu, Hunkeler, Ivezić, Jain, Jenness, Kanarek, Kendrew, Kern, Kerzendorf, Khvalko, King, Kirkby, Kulkarni, Kumar, Lee, Lenz, Littlefair, Ma, Macleod, Mastropietro, McCully, Montagnac, Morris, Mueller, Mumford, Muna, Murphy, Nelson, Nguyen, Ninan, Nöthe, Ogaz, Oh, Parejko, Parley, Pascual, Patil,
  Patil, Plunkett, Prochaska, Rastogi, Reddy~Janga, Sabater, Sakurikar, Seifert, Sherbert, Sherwood-Taylor, Shih, Sick, Silbiger, Singanamalla, Singer, Sladen, Sooley, Sornarajah, Streicher, Teuben, Thomas, Tremblay, Turner, Terrón, van Kerkwijk, de~la Vega, Watkins, Weaver, Whitmore, Woillez, Zabalza, \& {Astropy Contributors}}]{AstropyCollaboration2018}
{Astropy Collaboration}, Price-Whelan, A.~M., Sipőcz, B.~M., {et~al.} 2018, The Astronomical Journal, 156, 123, \dodoi{10.3847/1538-3881/aabc4f}

\bibitem[{{Astropy Collaboration} {et~al.}(2022){Astropy Collaboration}, Price-Whelan, Lim, Earl, Starkman, Bradley, Shupe, Patil, Corrales, Brasseur, Nöthe, Donath, Tollerud, Morris, Ginsburg, Vaher, Weaver, Tocknell, Jamieson, van Kerkwijk, Robitaille, Merry, Bachetti, Günther, Aldcroft, Alvarado-Montes, Archibald, Bódi, Bapat, Barentsen, Bazán, Biswas, Boquien, Burke, Cara, Cara, Conroy, Conseil, Craig, Cross, Cruz, D'Eugenio, Dencheva, Devillepoix, Dietrich, Eigenbrot, Erben, Ferreira, Foreman-Mackey, Fox, Freij, Garg, Geda, Glattly, Gondhalekar, Gordon, Grant, Greenfield, Groener, Guest, Gurovich, Handberg, Hart, Hatfield-Dodds, Homeier, Hosseinzadeh, Jenness, Jones, Joseph, Kalmbach, Karamehmetoglu, Kałuszyński, Kelley, Kern, Kerzendorf, Koch, Kulumani, Lee, Ly, Ma, MacBride, Maljaars, Muna, Murphy, Norman, O'Steen, Oman, Pacifici, Pascual, Pascual-Granado, Patil, Perren, Pickering, Rastogi, Roulston, Ryan, Rykoff, Sabater, Sakurikar, Salgado, Sanghi, Saunders, Savchenko, Schwardt, Seifert-Eckert,
  Shih, Jain, Shukla, Sick, Simpson, Singanamalla, Singer, Singhal, Sinha, Sipőcz, Spitler, Stansby, Streicher, Šumak, Swinbank, Taranu, Tewary, Tremblay, de~Val-Borro, Van~Kooten, Vasović, Verma, de~Miranda~Cardoso, Williams, Wilson, Winkel, Wood-Vasey, Xue, Yoachim, Zhang, Zonca, \& {Astropy Project Contributors}}]{AstropyCollaboration2022}
{Astropy Collaboration}, Price-Whelan, A.~M., Lim, P.~L., {et~al.} 2022, The Astrophysical Journal, 935, 167, \dodoi{10.3847/1538-4357/ac7c74}

\bibitem[{Banzatti {et~al.}(2023)Banzatti, Pontoppidan, Carr, Jellison, Pascucci, Najita, Romero-Mirza, Öberg, Kalyaan, Pinilla, Krijt, Long, Lambrechts, Rosotti, Herczeg, Salyk, Zhang, Bergin, Ballering, Meyer, Bruderer, \& {Jdiscs Collaboration}}]{Banzatti2023}
Banzatti, A., Pontoppidan, K.~M., Carr, J.~S., {et~al.} 2023, The Astrophysical Journal, 957, L22, \dodoi{10.3847/2041-8213/acf5ec}

\bibitem[{Barentine \& Lacy(2012)}]{Barentine2012}
Barentine, J.~C., \& Lacy, J.~H. 2012, The Astrophysical Journal, 757, 111, \dodoi{10.1088/0004-637X/757/2/111}

\bibitem[{Barr {et~al.}(2022{\natexlab{a}})Barr, Li, Boogert, Lee, DeWitt, \& Tielens}]{Barr2022a}
Barr, A.~G., Li, J., Boogert, A., {et~al.} 2022{\natexlab{a}}, Astronomy and Astrophysics, 666, A26, \dodoi{10.1051/0004-6361/202143003}

\bibitem[{Barr {et~al.}(2018)Barr, Boogert, DeWitt, Montiel, Richter, Indriolo, Neufeld, Pendleton, Chiar, Dungee, \& Tielens}]{Barr2018}
Barr, A.~G., Boogert, A., DeWitt, C.~N., {et~al.} 2018, The Astrophysical Journal, 868, L2, \dodoi{10.3847/2041-8213/aaeb23}

\bibitem[{Barr {et~al.}(2020)Barr, Boogert, DeWitt, Montiel, Richter, Lacy, Neufeld, Indriolo, Pendleton, Chiar, \& Tielens}]{Barr2020}
---. 2020, The Astrophysical Journal, 900, 104, \dodoi{10.3847/1538-4357/abab05}

\bibitem[{Barr {et~al.}(2022{\natexlab{b}})Barr, Boogert, Li, DeWitt, Montiel, Richter, Indriolo, Pendleton, Chiar, \& Tielens}]{Barr2022}
Barr, A.~G., Boogert, A., Li, J., {et~al.} 2022{\natexlab{b}}, The Astrophysical Journal, 935, 165, \dodoi{10.3847/1538-4357/ac74b8}

\bibitem[{Bast {et~al.}(2013)Bast, Lahuis, van Dishoeck, \& Tielens}]{Bast2013}
Bast, J.~E., Lahuis, F., van Dishoeck, E.~F., \& Tielens, A. G. G.~M. 2013, Astronomy and Astrophysics, 551, A118, \dodoi{10.1051/0004-6361/201219908}

\bibitem[{Beltrán \& Rivilla(2018)}]{Beltran2018}
Beltrán, M.~T., \& Rivilla, V.~M. 2018, in {ASP} {Conference} {Series}, Vol. 517 (Astronomical Society of the Pacific), 249, \dodoi{10.48550/arXiv.1806.08137}

\bibitem[{Bennett {et~al.}(2007)Bennett, Chen, Sun, Chang, \& Kaiser}]{Bennett2007}
Bennett, C.~J., Chen, S.-H., Sun, B.-J., Chang, A. H.~H., \& Kaiser, R.~I. 2007, The Astrophysical Journal, 660, 1588, \dodoi{10.1086/511296}

\bibitem[{Beuther {et~al.}(2012)Beuther, Linz, \& Henning}]{Beuther2012}
Beuther, H., Linz, H., \& Henning, T. 2012, Astronomy and Astrophysics, 543, A88, \dodoi{10.1051/0004-6361/201219128}

\bibitem[{Beuther {et~al.}(2013)Beuther, Linz, \& Henning}]{Beuther2013}
---. 2013, Astronomy and Astrophysics, 558, A81, \dodoi{10.1051/0004-6361/201321498}

\bibitem[{Beuther {et~al.}(2017)Beuther, Linz, Henning, Feng, \& Teague}]{Beuther2017}
Beuther, H., Linz, H., Henning, T., Feng, S., \& Teague, R. 2017, Astronomy and Astrophysics, 605, A61, \dodoi{10.1051/0004-6361/201730575}

\bibitem[{Bisschop {et~al.}(2007)Bisschop, Jørgensen, van Dishoeck, \& de~Wachter}]{Bisschop2007}
Bisschop, S.~E., Jørgensen, J.~K., van Dishoeck, E.~F., \& de~Wachter, E. B.~M. 2007, Astronomy and Astrophysics, 465, 913, \dodoi{10.1051/0004-6361:20065963}

\bibitem[{Blake {et~al.}(1987)Blake, Sutton, Masson, \& Phillips}]{Blake1987}
Blake, G.~A., Sutton, E.~C., Masson, C.~R., \& Phillips, T.~G. 1987, The Astrophysical Journal, 315, 621, \dodoi{10.1086/165165}

\bibitem[{Boley {et~al.}(2013)Boley, Linz, van Boekel, Henning, Feldt, Kaper, Leinert, Müller, Pascucci, Robberto, Stecklum, Waters, \& Zinnecker}]{Boley2013}
Boley, P.~A., Linz, H., van Boekel, R., {et~al.} 2013, Astronomy and Astrophysics, 558, A24, \dodoi{10.1051/0004-6361/201321539}

\bibitem[{Boogert {et~al.}(2008)Boogert, Pontoppidan, Knez, Lahuis, Kessler-Silacci, van Dishoeck, Blake, Augereau, Bisschop, Bottinelli, Brooke, Brown, Crapsi, Evans, Fraser, Geers, Huard, Jørgensen, Öberg, Allen, Harvey, Koerner, Mundy, Padgett, Sargent, \& Stapelfeldt}]{Boogert2008}
Boogert, A. C.~A., Pontoppidan, K.~M., Knez, C., {et~al.} 2008, The Astrophysical Journal, 678, 985, \dodoi{10.1086/533425}

\bibitem[{Boonman \& van Dishoeck(2003)}]{Boonman2003}
Boonman, A. M.~S., \& van Dishoeck, E.~F. 2003, Astronomy and Astrophysics, 403, 1003, \dodoi{10.1051/0004-6361:20030364}

\bibitem[{Boonman {et~al.}(2003)Boonman, van Dishoeck, Lahuis, \& Doty}]{Boonman2003a}
Boonman, A. M.~S., van Dishoeck, E.~F., Lahuis, F., \& Doty, S.~D. 2003, Astronomy and Astrophysics, 399, 1063, \dodoi{10.1051/0004-6361:20021868}

\bibitem[{Booth {et~al.}(2023)Booth, Law, Temmink, Leemker, \& Macías}]{Booth2023}
Booth, A.~S., Law, C.~J., Temmink, M., Leemker, M., \& Macías, E. 2023, Astronomy and Astrophysics, 678, A146, \dodoi{10.1051/0004-6361/202346974}

\bibitem[{Booth {et~al.}(2021)Booth, Walsh, Terwisscha~van Scheltinga, van Dishoeck, Ilee, Hogerheijde, Kama, \& Nomura}]{Booth2021}
Booth, A.~S., Walsh, C., Terwisscha~van Scheltinga, J., {et~al.} 2021, Nature Astronomy, 5, 684, \dodoi{10.1038/s41550-021-01352-w}

\bibitem[{Bosman {et~al.}(2022)Bosman, Bergin, Calahan, \& Duval}]{Bosman2022}
Bosman, A.~D., Bergin, E.~A., Calahan, J., \& Duval, S.~E. 2022, The Astrophysical Journal, 930, L26, \dodoi{10.3847/2041-8213/ac66ce}

\bibitem[{Boyer {et~al.}(2016)Boyer, Rivas, Tran, Verish, \& Arumainayagam}]{Boyer2016}
Boyer, M.~C., Rivas, N., Tran, A.~A., Verish, C.~A., \& Arumainayagam, C.~R. 2016, Surface Science, 652, 26, \dodoi{10.1016/j.susc.2016.03.012}

\bibitem[{Brauer {et~al.}(2012)Brauer, Sung, Pearson, Brown, \& Xu}]{Brauer2012}
Brauer, C.~S., Sung, K., Pearson, J.~C., Brown, L.~R., \& Xu, L.-H. 2012, Journal of Quantitative Spectroscopy and Radiative Transfer, 113, 128, \dodoi{10.1016/j.jqsrt.2011.09.012}

\bibitem[{Campbell \& Thompson(1984)}]{Campbell1984}
Campbell, B., \& Thompson, R.~I. 1984, The Astrophysical Journal, 279, 650, \dodoi{10.1086/161928}

\bibitem[{Carr \& Najita(2008)}]{Carr2008}
Carr, J.~S., \& Najita, J.~R. 2008, Science, 319, 1504, \dodoi{10.1126/science.1153807}

\bibitem[{Catone {et~al.}(2021)Catone, Satta, Castrovilli, Bolognesi, Avaldi, \& Cartoni}]{Catone2021}
Catone, D., Satta, M., Castrovilli, M.~C., {et~al.} 2021, Chemical Physics Letters, 771, 138467, \dodoi{10.1016/j.cplett.2021.138467}

\bibitem[{Ceccarelli(2004)}]{Ceccarelli2004}
Ceccarelli, C. 2004, 323, 195.
\newblock \url{https://ui.adsabs.harvard.edu/abs/2004ASPC..323..195C}

\bibitem[{Cesaroni(2005)}]{Cesaroni2005}
Cesaroni, R. 2005, in Proceedings {IAU} {Symposium}, Vol. 227 (International Astronomical Union), 59--69, \dodoi{10.1017/S1743921305004369}

\bibitem[{Charnley {et~al.}(1992)Charnley, Tielens, \& Millar}]{Charnley1992}
Charnley, S.~B., Tielens, A. G. G.~M., \& Millar, T.~J. 1992, The Astrophysical Journal, 399, L71, \dodoi{10.1086/186609}

\bibitem[{Chen {et~al.}(2024)Chen, Rocha, van Dishoeck, van Gelder, Nazari, Slavicinska, Francis, Tabone, Ressler, Klaassen, Beuther, Boogert, Gieser, Kavanagh, Perotti, Le~Gouellec, Majumdar, Güdel, \& Henning}]{Chen2024}
Chen, Y., Rocha, W. R.~M., van Dishoeck, E.~F., {et~al.} 2024, Astronomy and Astrophysics, 690, A205, \dodoi{10.1051/0004-6361/202450706}

\bibitem[{Cuadrado {et~al.}(2017)Cuadrado, Goicoechea, Cernicharo, Fuente, Pety, \& Tercero}]{Cuadrado2017}
Cuadrado, S., Goicoechea, J.~R., Cernicharo, J., {et~al.} 2017, Astronomy and Astrophysics, 603, A124, \dodoi{10.1051/0004-6361/201730459}

\bibitem[{Davis {et~al.}(1998)Davis, Moriarty-Schieven, Eislöffel, Hoare, \& Ray}]{Davis1998}
Davis, C.~J., Moriarty-Schieven, G., Eislöffel, J., Hoare, M.~G., \& Ray, T.~P. 1998, The Astronomical Journal, 115, 1118, \dodoi{10.1086/300259}

\bibitem[{De~Buizer(2006)}]{DeBuizer2006}
De~Buizer, J.~M. 2006, The Astrophysical Journal, 642, L57, \dodoi{10.1086/504291}

\bibitem[{De~Buizer \& Minier(2005)}]{DeBuizer2005}
De~Buizer, J.~M., \& Minier, V. 2005, The Astrophysical Journal, 628, L151, \dodoi{10.1086/432835}

\bibitem[{De~Buizer {et~al.}(2017)De~Buizer, Liu, Tan, Zhang, Beltrán, Shuping, Staff, Tanaka, \& Whitney}]{Debuizer2017}
De~Buizer, J.~M., Liu, M., Tan, J.~C., {et~al.} 2017, The Astrophysical Journal, 843, 33, \dodoi{10.3847/1538-4357/aa74c8}

\bibitem[{de~Graauw {et~al.}(1996)de~Graauw, Haser, Beintema, Roelfsema, van Agthoven, Barl, Bauer, Bekenkamp, Boonstra, Boxhoorn, Cote, de~Groene, van Dijkhuizen, Drapatz, Evers, Feuchtgruber, Frericks, Genzel, Haerendel, Heras, van~der Hucht, van~der Hulst, Huygen, Jacobs, Jakob, Kamperman, Katterloher, Kester, Kunze, Kussendrager, Lahuis, Lamers, Leech, van~der Lei, van~der Linden, Luinge, Lutz, Melzner, Morris, van Nguyen, Ploeger, Price, Salama, Schaeidt, Sijm, Smoorenburg, Spakman, Spoon, Steinmayer, Stoecker, Valentijn, Vandenbussche, Visser, Waelkens, Waters, Wensink, Wesselius, Wiezorrek, Wieprecht, Wijnbergen, Wildeman, \& Young}]{DeGraauw1996}
de~Graauw, T., Haser, L.~N., Beintema, D.~A., {et~al.} 1996, Astronomy and Astrophysics, 315, L49.
\newblock \url{https://ui.adsabs.harvard.edu/abs/1996A&A...315L..49D}

\bibitem[{de~Wit {et~al.}(2010)de~Wit, Hoare, Oudmaijer, \& Lumsden}]{deWit2010}
de~Wit, W.~J., Hoare, M.~G., Oudmaijer, R.~D., \& Lumsden, S.~L. 2010, Astronomy and Astrophysics, 515, A45, \dodoi{10.1051/0004-6361/200913209}

\bibitem[{DeWitt {et~al.}(2023)DeWitt, Montiel, \& Rashman}]{DeWitt2023}
DeWitt, C., Montiel, E., \& Rashman, M. 2023, {EXES} {Handbook} for {Archive} {Users}, {SOFIA} - {Stratospheric} {Observatory} for {Infrared} {Astronomy}, {Rev1}.0.
\newblock \url{https://irsa.ipac.caltech.edu/data/SOFIA/docs/instruments/handbooks/EXES_Handbook_for_Archive_Users_Ver1.0.pdf}

\bibitem[{Draine(2011)}]{Draine2011}
Draine, B.~T. 2011, Physics of the {Interstellar} and {Intergalactic} {Medium}.
\newblock \url{https://ui.adsabs.harvard.edu/abs/2011piim.book.....D}

\bibitem[{Dullemond {et~al.}(2007)Dullemond, Hollenbach, Kamp, \& D'Alessio}]{Dullemond2007}
Dullemond, C.~P., Hollenbach, D., Kamp, I., \& D'Alessio, P. 2007, Models of the {Structure} and {Evolution} of {Protoplanetary} {Disks} (eprint: arXiv:astro-ph/0602619), \dodoi{10.48550/arXiv.astro-ph/0602619}

\bibitem[{Dungee {et~al.}(2018)Dungee, Boogert, DeWitt, Montiel, Richter, Barr, Blake, Charnley, Indriolo, Karska, Neufeld, Smith, \& Tielens}]{Dungee2018}
Dungee, R., Boogert, A., DeWitt, C.~N., {et~al.} 2018, The Astrophysical Journal, 868, L10, \dodoi{10.3847/2041-8213/aaeda9}

\bibitem[{Friberg {et~al.}(1988)Friberg, Madden, Hjalmarson, \& Irvine}]{Friberg1988}
Friberg, P., Madden, S.~C., Hjalmarson, A., \& Irvine, W.~M. 1988, Astronomy and Astrophysics, 195, 281.
\newblock \url{https://ui.adsabs.harvard.edu/abs/1988A&A...195..281F}

\bibitem[{Fuchs {et~al.}(2009)Fuchs, Cuppen, Ioppolo, Romanzin, Bisschop, Andersson, van Dishoeck, \& Linnartz}]{Fuchs2009}
Fuchs, G.~W., Cuppen, H.~M., Ioppolo, S., {et~al.} 2009, Astronomy and Astrophysics, 505, 629, \dodoi{10.1051/0004-6361/200810784}

\bibitem[{Garrod {et~al.}(2006)Garrod, Park, Caselli, \& Herbst}]{Garrod2006}
Garrod, R., Park, I.~H., Caselli, P., \& Herbst, E. 2006, Faraday Discussions, 133, 51, \dodoi{10.1039/b516202e}

\bibitem[{Garrod {et~al.}(2008)Garrod, Widicus~Weaver, \& Herbst}]{Garrod2008}
Garrod, R.~T., Widicus~Weaver, S.~L., \& Herbst, E. 2008, The Astrophysical Journal, 682, 283, \dodoi{10.1086/588035}

\bibitem[{Ginsburg {et~al.}(2019)Ginsburg, Sipőcz, Brasseur, Cowperthwaite, Craig, Deil, Guillochon, Guzman, Liedtke, Lian~Lim, Lockhart, Mommert, Morris, Norman, Parikh, Persson, Robitaille, Segovia, Singer, Tollerud, de~Val-Borro, Valtchanov, Woillez, {Astroquery Collaboration}, \& {a subset of astropy Collaboration}}]{Ginsburg2019a}
Ginsburg, A., Sipőcz, B.~M., Brasseur, C.~E., {et~al.} 2019, The Astronomical Journal, 157, 98, \dodoi{10.3847/1538-3881/aafc33}

\bibitem[{Goddi {et~al.}(2015)Goddi, Zhang, \& Moscadelli}]{Goddi2015}
Goddi, C., Zhang, Q., \& Moscadelli, L. 2015, Astronomy and Astrophysics, 573, A108, \dodoi{10.1051/0004-6361/201424832}

\bibitem[{Goldsmith \& Langer(1999)}]{Goldsmith1999}
Goldsmith, P.~F., \& Langer, W.~D. 1999, The Astrophysical Journal, 517, 209, \dodoi{10.1086/307195}

\bibitem[{Gordon {et~al.}(2022)Gordon, Rothman, Hargreaves, Hashemi, Karlovets, Skinner, Conway, Hill, Kochanov, Tan, Wcisło, Finenko, Nelson, Bernath, Birk, Boudon, Campargue, Chance, Coustenis, Drouin, Flaud, Gamache, Hodges, Jacquemart, Mlawer, Nikitin, Perevalov, Rotger, Tennyson, Toon, Tran, Tyuterev, Adkins, Baker, Barbe, Canè, Császár, Dudaryonok, Egorov, Fleisher, Fleurbaey, Foltynowicz, Furtenbacher, Harrison, Hartmann, Horneman, Huang, Karman, Karns, Kassi, Kleiner, Kofman, Kwabia–Tchana, Lavrentieva, Lee, Long, Lukashevskaya, Lyulin, Makhnev, Matt, Massie, Melosso, Mikhailenko, Mondelain, Müller, Naumenko, Perrin, Polyansky, Raddaoui, Raston, Reed, Rey, Richard, Tóbiás, Sadiek, Schwenke, Starikova, Sung, Tamassia, Tashkun, Vander~Auwera, Vasilenko, Vigasin, Villanueva, Vispoel, Wagner, Yachmenev, \& Yurchenko}]{Gordon2022}
Gordon, I.~E., Rothman, L.~S., Hargreaves, R.~J., {et~al.} 2022, Journal of Quantitative Spectroscopy and Radiative Transfer, 277, 107949, \dodoi{10.1016/j.jqsrt.2021.107949}

\bibitem[{Goto {et~al.}(2019)Goto, Geballe, Harju, Caselli, Sipilä, Menten, \& Usuda}]{Goto2019}
Goto, M., Geballe, T.~R., Harju, J., {et~al.} 2019, Astronomy and Astrophysics, 632, A29, \dodoi{10.1051/0004-6361/201936119}

\bibitem[{Goto {et~al.}(2015)Goto, Geballe, \& Usuda}]{Goto2015}
Goto, M., Geballe, T.~R., \& Usuda, T. 2015, The Astrophysical Journal, 806, 57, \dodoi{10.1088/0004-637X/806/1/57}

\bibitem[{Guzmán {et~al.}(2013)Guzmán, Goicoechea, Pety, Gratier, Gerin, Roueff, Le~Petit, Le~Bourlot, \& Faure}]{Guzman2013}
Guzmán, V.~V., Goicoechea, J.~R., Pety, J., {et~al.} 2013, Astronomy and Astrophysics, 560, A73, \dodoi{10.1051/0004-6361/201322460}

\bibitem[{Habing {et~al.}(1972)Habing, Israel, \& de~Jong}]{Habing1972}
Habing, H.~J., Israel, F.~P., \& de~Jong, T. 1972, Astronomy and Astrophysics, 17, 329.
\newblock \url{https://ui.adsabs.harvard.edu/abs/1972A&A....17..329H}

\bibitem[{Hackwell {et~al.}(1982)Hackwell, Grasdalen, \& Gehrz}]{Hackwell1982}
Hackwell, J.~A., Grasdalen, G.~L., \& Gehrz, R.~D. 1982, The Astrophysical Journal, 252, 250, \dodoi{10.1086/159552}

\bibitem[{Harris {et~al.}(2020)Harris, Millman, van~der Walt, Gommers, Virtanen, Cournapeau, Wieser, Taylor, Berg, Smith, Kern, Picus, Hoyer, van Kerkwijk, Brett, Haldane, del Río, Wiebe, Peterson, Gérard-Marchant, Sheppard, Reddy, Weckesser, Abbasi, Gohlke, \& Oliphant}]{Harris2020}
Harris, C.~R., Millman, K.~J., van~der Walt, S.~J., {et~al.} 2020, Nature, 585, 357, \dodoi{10.1038/s41586-020-2649-2}

\bibitem[{Herter {et~al.}(2013)Herter, Vacca, Adams, Keller, Schoenwald, Hirsch, Wang, De~Buizer, Helton, \& Llorens}]{Herter2013}
Herter, T.~L., Vacca, W.~D., Adams, J.~D., {et~al.} 2013, Publications of the Astronomical Society of the Pacific, 125, 1393, \dodoi{10.1086/674144}

\bibitem[{Hillman {et~al.}(1991)Hillman, Jennings, Halsey, Nadler, \& Blass}]{Hillman1991}
Hillman, J.~J., Jennings, D.~E., Halsey, G.~W., Nadler, S., \& Blass, W.~E. 1991, Journal of Molecular Spectroscopy, 146, 389, \dodoi{10.1016/0022-2852(91)90014-2}

\bibitem[{Humire {et~al.}(2025)Humire, Dey, Ronconi, Sasse, Cid~Fernandes, Martín, Donevski, Małek, Fernández-Ontiveros, Song, Hamed, Mangum, Henkel, Rivilla, Colzi, Harada, Demarco, Goyal, Meier, Panda, Krabbe, Yan, Lopes, Sakamoto, Muller, Tanaka, Yoshimura, Nakanishi, Kanaan, Ribeiro, Schoenell, \& Mendes~de Oliveira}]{Humire2025}
Humire, P.~K., Dey, S., Ronconi, T., {et~al.} 2025, Spatially-resolved spectro-photometric {SED} {Modeling} of {NGC} 253's {Central} {Molecular} {Zone} {I}. {Studying} the star formation in extragalactic giant molecular clouds, \dodoi{10.48550/arXiv.2501.15082}

\bibitem[{Hunter(2007)}]{Hunter2007}
Hunter, J.~D. 2007, Computing in Science and Engineering, 9, 90, \dodoi{10.1109/MCSE.2007.55}

\bibitem[{Ilee {et~al.}(2021)Ilee, Walsh, Booth, Aikawa, Andrews, Bae, Bergin, Bergner, Bosman, Cataldi, Cleeves, Czekala, Guzmán, Huang, Law, Le~Gal, Loomis, Ménard, Nomura, Öberg, Qi, Schwarz, Teague, Tsukagoshi, Wilner, Yamato, \& Zhang}]{Ilee2021}
Ilee, J.~D., Walsh, C., Booth, A.~S., {et~al.} 2021, The Astrophysical Journal Supplement Series, 257, 9, \dodoi{10.3847/1538-4365/ac1441}

\bibitem[{Indriolo {et~al.}(2015)Indriolo, Neufeld, DeWitt, Richter, Boogert, Harper, Jaffe, Kulas, McKelvey, Ryde, \& Vacca}]{Indriolo2015}
Indriolo, N., Neufeld, D.~A., DeWitt, C.~N., {et~al.} 2015, The Astrophysical Journal, 802, L14, \dodoi{10.1088/2041-8205/802/2/L14}

\bibitem[{Indriolo {et~al.}(2020)Indriolo, Neufeld, Barr, Boogert, DeWitt, Karska, Montiel, Richter, \& Tielens}]{Indriolo2020}
Indriolo, N., Neufeld, D.~A., Barr, A.~G., {et~al.} 2020, The Astrophysical Journal, 894, 107, \dodoi{10.3847/1538-4357/ab88a1}

\bibitem[{Jin \& Garrod(2020)}]{Jin2020}
Jin, M., \& Garrod, R.~T. 2020, The Astrophysical Journal Supplement Series, 249, 26, \dodoi{10.3847/1538-4365/ab9ec8}

\bibitem[{Jones {et~al.}(2023)Jones, Álvarez Márquez, Sloan, Kavanagh, Argyriou, Law, Labiano, Patapis, Mueller, Larson, Bright, Klaassen, Fox, Gasman, Geers, Glauser, Guillard, Nayak, Noriega-Crespo, Ressler, Sargent, Temim, Vandenbussche, \& García~Marín}]{Jones2023}
Jones, O.~C., Álvarez Márquez, J., Sloan, G.~C., {et~al.} 2023, Monthly Notices of the Royal Astronomical Society, 523, 2519, \dodoi{10.1093/mnras/stad1609}

\bibitem[{Kabbadj {et~al.}(1991)Kabbadj, Herman, Di~Lonardo, Fusina, \& Johns}]{Kabbadj1991}
Kabbadj, Y., Herman, M., Di~Lonardo, G., Fusina, L., \& Johns, J. W.~C. 1991, Journal of Molecular Spectroscopy, 150, 535, \dodoi{10.1016/0022-2852(91)90248-9}

\bibitem[{Kaiser \& Roessler(1998)}]{Kaiser1998}
Kaiser, R.~I., \& Roessler, K. 1998, The Astrophysical Journal, 503, 959, \dodoi{10.1086/306001}

\bibitem[{Kameya {et~al.}(1989)Kameya, Hasegawa, Hirano, Takakubo, \& Seki}]{Kameya1989}
Kameya, O., Hasegawa, T.~I., Hirano, N., Takakubo, K., \& Seki, M. 1989, The Astrophysical Journal, 339, 222, \dodoi{10.1086/167289}

\bibitem[{Kamp {et~al.}(2018)Kamp, Antonellini, Carmona, Ilee, \& Rab}]{Kamp2018}
Kamp, I., Antonellini, S., Carmona, A., Ilee, J., \& Rab, C. 2018, Multi-wavelength observations of planet forming disks: {Constraints} on planet formation processes (eprint: arXiv:1712.00303), \dodoi{10.48550/arXiv.1712.00303}

\bibitem[{Keane {et~al.}(2001)Keane, Boonman, Tielens, \& van Dishoeck}]{Keane2001}
Keane, J.~V., Boonman, A. M.~S., Tielens, A. G. G.~M., \& van Dishoeck, E.~F. 2001, Astronomy and Astrophysics, 376, L5, \dodoi{10.1051/0004-6361:20011008}

\bibitem[{Knez {et~al.}(2009)Knez, Lacy, Evans, van Dishoeck, \& Richter}]{Knez2009}
Knez, C., Lacy, J.~H., Evans, II, N.~J., van Dishoeck, E.~F., \& Richter, M.~J. 2009, The Astrophysical Journal, 696, 471, \dodoi{10.1088/0004-637X/696/1/471}

\bibitem[{Kochanov {et~al.}(2016)Kochanov, Gordon, Rothman, Wcisło, Hill, \& Wilzewski}]{Kochanov2016}
Kochanov, R.~V., Gordon, I.~E., Rothman, L.~S., {et~al.} 2016, Journal of Quantitative Spectroscopy and Radiative Transfer, 177, 15, \dodoi{10.1016/j.jqsrt.2016.03.005}

\bibitem[{Kurtz {et~al.}(2000)Kurtz, Cesaroni, Churchwell, Hofner, \& Walmsley}]{Kurtz2000}
Kurtz, S., Cesaroni, R., Churchwell, E., Hofner, P., \& Walmsley, C.~M. 2000, in Protostars and {Planets} {IV}, 299--326.
\newblock \url{https://ui.adsabs.harvard.edu/abs/2000prpl.conf..299K}

\bibitem[{Lacy {et~al.}(1989)Lacy, Evans, Achtermann, Bruce, Arens, \& Carr}]{Lacy1989}
Lacy, J.~H., Evans, II, N.~J., Achtermann, J.~M., {et~al.} 1989, The Astrophysical Journal, 342, L43, \dodoi{10.1086/185480}

\bibitem[{Lacy {et~al.}(2002)Lacy, Richter, Greathouse, Jaffe, \& Zhu}]{Lacy2002}
Lacy, J.~H., Richter, M.~J., Greathouse, T.~K., Jaffe, D.~T., \& Zhu, Q. 2002, Publications of the Astronomical Society of the Pacific, 114, 153, \dodoi{10.1086/338730}

\bibitem[{Lahuis \& van Dishoeck(2000)}]{Lahuis2000}
Lahuis, F., \& van Dishoeck, E.~F. 2000, Astronomy and Astrophysics, 355, 699.
\newblock \url{https://ui.adsabs.harvard.edu/abs/2000A&A...355..699L}

\bibitem[{Lees \& Baker(1968)}]{Lees1968}
Lees, R.~M., \& Baker, J.~G. 1968, Journal of Chemical Physics, 48, 5299, \dodoi{10.1063/1.1668221}

\bibitem[{Lees {et~al.}(2020)Lees, Xu, \& Billinghurst}]{Lees2020}
Lees, R.~M., Xu, L.-H., \& Billinghurst, B.~E. 2020, Journal of Molecular Structure, 1209, 127960, \dodoi{10.1016/j.molstruc.2020.127960}

\bibitem[{Li {et~al.}(2022)Li, Boogert, Barr, \& Tielens}]{Li2022}
Li, J., Boogert, A., Barr, A.~G., \& Tielens, A. G. G.~M. 2022, The Astrophysical Journal, 935, 161, \dodoi{10.3847/1538-4357/ac7ce7}

\bibitem[{Li {et~al.}(2023)Li, Boogert, Barr, DeWitt, Rashman, Neufeld, Indriolo, Pendleton, Montiel, Richter, Chiar, \& Tielens}]{Li2023}
Li, J., Boogert, A., Barr, A.~G., {et~al.} 2023, The Astrophysical Journal, 953, 103, \dodoi{10.3847/1538-4357/ace16e}

\bibitem[{Lord(1992)}]{Lord1992}
Lord, S.~D. 1992, Infrared Radiation

\bibitem[{MacDonald {et~al.}(1996)MacDonald, Gibb, Habing, \& Millar}]{Macdonald1996}
MacDonald, G.~H., Gibb, A.~G., Habing, R.~J., \& Millar, T.~J. 1996, Astronomy and Astrophysics Supplement Series, 119, 333.
\newblock \url{https://ui.adsabs.harvard.edu/abs/1996A&AS..119..333M}

\bibitem[{Maret {et~al.}(2005)Maret, Ceccarelli, Tielens, Caux, Lefloch, Faure, Castets, \& Flower}]{Maret2005}
Maret, S., Ceccarelli, C., Tielens, A. G. G.~M., {et~al.} 2005, Astronomy and Astrophysics, 442, 527, \dodoi{10.1051/0004-6361:20052899}

\bibitem[{Martin(1973)}]{Martin1973}
Martin, A. H.~M. 1973, Monthly Notices of the Royal Astronomical Society, 163, 141, \dodoi{10.1093/mnras/163.2.141}

\bibitem[{McClure {et~al.}(2023)McClure, Rocha, Pontoppidan, Crouzet, Chu, Dartois, Lamberts, Noble, Pendleton, Perotti, Qasim, Rachid, Smith, Sun, Beck, Boogert, Brown, Caselli, Charnley, Cuppen, Dickinson, Drozdovskaya, Egami, Erkal, Fraser, Garrod, Harsono, Ioppolo, Jiménez-Serra, Jin, Jørgensen, Kristensen, Lis, McCoustra, McGuire, Melnick, Ã-berg, Palumbo, Shimonishi, Sturm, van Dishoeck, \& Linnartz}]{McClure2023}
McClure, M.~K., Rocha, W. R.~M., Pontoppidan, K.~M., {et~al.} 2023, Nature Astronomy, 7, 431, \dodoi{10.1038/s41550-022-01875-w}

\bibitem[{Mendoza {et~al.}(2018)Mendoza, Bronfman, Duronea, Lépine, Finger, Merello, Hervías-Caimapo, Gama, Reyes, \& Åke Nyman}]{Mendoza2018}
Mendoza, E., Bronfman, L., Duronea, N.~U., {et~al.} 2018, The Astrophysical Journal, 853, 152, \dodoi{10.3847/1538-4357/aaa1ec}

\bibitem[{Minier {et~al.}(2000)Minier, Booth, \& Conway}]{Minier2000}
Minier, V., Booth, R.~S., \& Conway, J.~E. 2000, Astronomy and Astrophysics, 362, 1093.
\newblock \url{https://ui.adsabs.harvard.edu/abs/2000A&A...362.1093M}

\bibitem[{Moore \& Hudson(1998)}]{Moore1998}
Moore, M.~H., \& Hudson, R.~L. 1998, Icarus, 135, 518, \dodoi{10.1006/icar.1998.5996}

\bibitem[{Moruzzi {et~al.}(1995)Moruzzi, Winnewisser, Winnewisser, Mukhopadhyay, \& Strumia}]{Moruzzi1995}
Moruzzi, G., Winnewisser, B.~P., Winnewisser, M., Mukhopadhyay, I., \& Strumia, F. 1995, in Millimeter and {Submillimeter} {Waves} {II}, Vol. 2558, 285--292, \dodoi{10.1117/12.224249}

\bibitem[{Moscadelli \& Goddi(2014)}]{Moscadelli2014}
Moscadelli, L., \& Goddi, C. 2014, Astronomy and Astrophysics, 566, A150, \dodoi{10.1051/0004-6361/201423420}

\bibitem[{Moscadelli {et~al.}(2025)Moscadelli, Goddi, Hirota, \& Sanna}]{Moscadelli2025}
Moscadelli, L., Goddi, C., Hirota, T., \& Sanna, A. 2025, Astronomy and Astrophysics, 696, A47, \dodoi{10.1051/0004-6361/202553842}

\bibitem[{Moscadelli {et~al.}(2009)Moscadelli, Reid, Menten, Brunthaler, Zheng, \& Xu}]{Moscadelli2009}
Moscadelli, L., Reid, M.~J., Menten, K.~M., {et~al.} 2009, The Astrophysical Journal, 693, 406, \dodoi{10.1088/0004-637X/693/1/406}

\bibitem[{Nazari {et~al.}(2023)Nazari, Tabone, \& Rosotti}]{Nazari2023}
Nazari, P., Tabone, B., \& Rosotti, G.~P. 2023, Astronomy and Astrophysics, 671, A107, \dodoi{10.1051/0004-6361/202244801}

\bibitem[{Nickerson {et~al.}(2021)Nickerson, Rangwala, Colgan, DeWitt, Huang, Acharyya, Drozdovskaya, Fortenberry, Herbst, \& Lee}]{Nickerson2021}
Nickerson, S., Rangwala, N., Colgan, S. W.~J., {et~al.} 2021, The Astrophysical Journal, 907, 51, \dodoi{10.3847/1538-4357/abca36}

\bibitem[{Nickerson {et~al.}(2023)Nickerson, Rangwala, Colgan, DeWitt, Monzon, Huang, Acharyya, Drozdovskaya, Fortenberry, Herbst, \& Lee}]{Nickerson2023}
---. 2023, The Astrophysical Journal, 945, 26, \dodoi{10.3847/1538-4357/aca6e8}

\bibitem[{Pestalozzi {et~al.}(2004)Pestalozzi, Elitzur, Conway, \& Booth}]{Pestalozzi2004}
Pestalozzi, M.~R., Elitzur, M., Conway, J.~E., \& Booth, R.~S. 2004, The Astrophysical Journal, 603, L113, \dodoi{10.1086/383127}

\bibitem[{Punanova {et~al.}(2022)Punanova, Vasyunin, Caselli, Howard, Spezzano, Shirley, Scibelli, \& Harju}]{Punanova2022}
Punanova, A., Vasyunin, A., Caselli, P., {et~al.} 2022, The Astrophysical Journal, 927, 213, \dodoi{10.3847/1538-4357/ac4e7d}

\bibitem[{Rangwala {et~al.}(2018)Rangwala, Colgan, Le~Gal, Acharyya, Huang, Lee, Herbst, deWitt, Richter, Boogert, \& McKelvey}]{Rangwala2018}
Rangwala, N., Colgan, S. W.~J., Le~Gal, R., {et~al.} 2018, The Astrophysical Journal, 856, 9, \dodoi{10.3847/1538-4357/aaab66}

\bibitem[{Richter {et~al.}(2018)Richter, Dewitt, McKelvey, Montiel, McMurray, \& Case}]{Richter2018}
Richter, M.~J., Dewitt, C.~N., McKelvey, M., {et~al.} 2018, Journal of Astronomical Instrumentation, 7, 1840013, \dodoi{10.1142/S2251171718400135}

\bibitem[{Rieke {et~al.}(2015)Rieke, Ressler, Morrison, Bergeron, Bouchet, García-Marín, Greene, Regan, Sukhatme, \& Walker}]{Rieke2015}
Rieke, G.~H., Ressler, M.~E., Morrison, J.~E., {et~al.} 2015, Publications of the Astronomical Society of the Pacific, 127, 665, \dodoi{10.1086/682257}

\bibitem[{Sandell {et~al.}(2020)Sandell, Wright, Güsten, Wiesemeyer, Reyes, Mookerjea, \& Corder}]{Sandell2020}
Sandell, G., Wright, M., Güsten, R., {et~al.} 2020, The Astrophysical Journal, 904, 139, \dodoi{10.3847/1538-4357/abbf5b}

\bibitem[{Sandford {et~al.}(2020)Sandford, Nuevo, Bera, \& Lee}]{Sandford2020}
Sandford, S.~A., Nuevo, M., Bera, P.~P., \& Lee, T.~J. 2020, Chemical Reviews, 120, 4616, \dodoi{10.1021/acs.chemrev.9b00560}

\bibitem[{Scoville {et~al.}(1986)Scoville, Sargent, Sanders, Claussen, Masson, Lo, \& Phillips}]{Scoville1986}
Scoville, N.~Z., Sargent, A.~I., Sanders, D.~B., {et~al.} 1986, The Astrophysical Journal, 303, 416, \dodoi{10.1086/164086}

\bibitem[{Sharpless(1959)}]{Sharpless1959}
Sharpless, S. 1959, The Astrophysical Journal Supplement Series, 4, 257, \dodoi{10.1086/190049}

\bibitem[{Soma {et~al.}(2015)Soma, Sakai, Watanabe, \& Yamamoto}]{Soma2015}
Soma, T., Sakai, N., Watanabe, Y., \& Yamamoto, S. 2015, The Astrophysical Journal, 802, 74, \dodoi{10.1088/0004-637X/802/2/74}

\bibitem[{Surcis {et~al.}(2011)Surcis, Vlemmings, Torres, van Langevelde, \& Hutawarakorn~Kramer}]{Surcis2011}
Surcis, G., Vlemmings, W. H.~T., Torres, R.~M., van Langevelde, H.~J., \& Hutawarakorn~Kramer, B. 2011, Astronomy and Astrophysics, 533, A47, \dodoi{10.1051/0004-6361/201117108}

\bibitem[{Turner(1998)}]{Turner1998}
Turner, B.~E. 1998, The Astrophysical Journal, 501, 731, \dodoi{10.1086/305859}

\bibitem[{van~der Tak(2004)}]{VanDerTak2004}
van~der Tak, F. F.~S. 2004, in {IAU} {Symposium}, Vol. 221 (International Astronomical Union), 59, \dodoi{10.48550/arXiv.astro-ph/0309152}

\bibitem[{van~der Tak {et~al.}(2000)van~der Tak, van Dishoeck, \& Caselli}]{VanDerTak2000}
van~der Tak, F. F.~S., van Dishoeck, E.~F., \& Caselli, P. 2000, Astronomy and Astrophysics, 361, 327, \dodoi{10.48550/arXiv.astro-ph/0008010}

\bibitem[{van Dishoeck \& Blake(1998)}]{VanDishoeck1998}
van Dishoeck, E.~F., \& Blake, G.~A. 1998, Annual Review of Astronomy and Astrophysics, 36, 317, \dodoi{10.1146/annurev.astro.36.1.317}

\bibitem[{van Dishoeck {et~al.}(1995)van Dishoeck, Blake, Jansen, \& Groesbeck}]{VanDishoeck1995}
van Dishoeck, E.~F., Blake, G.~A., Jansen, D.~J., \& Groesbeck, T.~D. 1995, The Astrophysical Journal, 447, 760, \dodoi{10.1086/175915}

\bibitem[{Villanueva {et~al.}(2012)Villanueva, DiSanti, Mumma, \& Xu}]{Villanueva2012}
Villanueva, G.~L., DiSanti, M.~A., Mumma, M.~J., \& Xu, L.~H. 2012, The Astrophysical Journal, 747, 37, \dodoi{10.1088/0004-637X/747/1/37}

\bibitem[{Virtanen {et~al.}(2020)Virtanen, Gommers, Oliphant, Haberland, Reddy, Cournapeau, Burovski, Peterson, Weckesser, Bright, van~der Walt, Brett, Wilson, Millman, Mayorov, Nelson, Jones, Kern, Larson, Carey, Polat, Feng, Moore, VanderPlas, Laxalde, Perktold, Cimrman, Henriksen, Quintero, Harris, Archibald, Ribeiro, Pedregosa, van Mulbregt, \& {SciPy 1. 0 Contributors}}]{Virtanen2020}
Virtanen, P., Gommers, R., Oliphant, T.~E., {et~al.} 2020, Nature Methods, 17, 261, \dodoi{10.1038/s41592-019-0686-2}

\bibitem[{Walsh {et~al.}(2016)Walsh, Loomis, Öberg, Kama, van~'t Hoff, Millar, Aikawa, Herbst, Widicus~Weaver, \& Nomura}]{Walsh2016}
Walsh, C., Loomis, R.~A., Öberg, K.~I., {et~al.} 2016, The Astrophysical Journal, 823, L10, \dodoi{10.3847/2041-8205/823/1/L10}

\bibitem[{Watanabe \& Kouchi(2002)}]{Watanabe2002}
Watanabe, N., \& Kouchi, A. 2002, The Astrophysical Journal, 571, L173, \dodoi{10.1086/341412}

\bibitem[{Werner {et~al.}(1979)Werner, Becklin, Gatley, Matthews, Neugebauer, \& Wynn-Williams}]{Werner1979}
Werner, M.~W., Becklin, E.~E., Gatley, I., {et~al.} 1979, Monthly Notices of the Royal Astronomical Society, 188, 463, \dodoi{10.1093/mnras/188.3.463}

\bibitem[{Wiesemeyer {et~al.}(2004)Wiesemeyer, Thum, \& Walmsley}]{Wiesemeyer2004}
Wiesemeyer, H., Thum, C., \& Walmsley, C.~M. 2004, Astronomy and Astrophysics, 428, 479, \dodoi{10.1051/0004-6361:20040343}

\bibitem[{Williams \& Cieza(2011)}]{Williams2011}
Williams, J.~P., \& Cieza, L.~A. 2011, Annual Review of Astronomy and Astrophysics, 49, 67, \dodoi{10.1146/annurev-astro-081710-102548}

\bibitem[{Willner(1976)}]{Willner1976}
Willner, S.~P. 1976, The Astrophysical Journal, 206, 728, \dodoi{10.1086/154433}

\bibitem[{Wirström {et~al.}(2011)Wirström, Geppert, Hjalmarson, Persson, Black, Bergman, Millar, Hamberg, \& Vigren}]{Wirstrom2011}
Wirström, E.~S., Geppert, W.~D., Hjalmarson, A., {et~al.} 2011, Astronomy and Astrophysics, 533, A24, \dodoi{10.1051/0004-6361/201116525}

\bibitem[{Wright {et~al.}(2015)Wright, Wright, Goodson, Rieke, Aitink-Kroes, Amiaux, Aricha-Yanguas, Azzollini, Banks, Barrado-Navascues, Belenguer-Davila, Blommaert, Bouchet, Brandl, Colina, Detre, Diaz-Catala, Eccleston, Friedman, García-Marín, Güdel, Glasse, Glauser, Greene, Groezinger, Grundy, Hastings, Henning, Hofferbert, Hunter, Jessen, Justtanont, Karnik, Khorrami, Krause, Labiano, Lagage, Langer, Lemke, Lim, Lorenzo-Alvarez, Mazy, McGowan, Meixner, Morris, Morrison, Müller, rgaard Nielson, Olofsson, O'Sullivan, Pel, Penanen, Petach, Pye, Ray, Renotte, Renouf, Ressler, Samara-Ratna, Scheithauer, Schneider, Shaughnessy, Stevenson, Sukhatme, Swinyard, Sykes, Thatcher, Tikkanen, van Dishoeck, Waelkens, Walker, Wells, \& Zhender}]{Wright2015}
Wright, G.~S., Wright, D., Goodson, G.~B., {et~al.} 2015, Publications of the Astronomical Society of the Pacific, 127, 595, \dodoi{10.1086/682253}

\bibitem[{Wright {et~al.}(2023)Wright, Rieke, Glasse, Ressler, García~Marín, Aguilar, Alberts, Álvarez Márquez, Argyriou, Banks, Baudoz, Boccaletti, Bouchet, Bouwman, Brandl, Breda, Bright, Cale, Colina, Cossou, Coulais, Cracraft, De~Meester, Dicken, Engesser, Etxaluze, Fox, Friedman, Fu, Gasman, Gáspár, Gastaud, Geers, Glauser, Gordon, Greene, Greve, Grundy, Güdel, Guillard, Haderlein, Hashimoto, Henning, Hines, Holler, Detre, Jahromi, James, Jones, Justtanont, Kavanagh, Kendrew, Klaassen, Krause, Labiano, Lagage, Lambros, Larson, Law, Lee, Libralato, Lorenzo~Alverez, Meixner, Morrison, Mueller, Murray, Mycroft, Myers, Nayak, Naylor, Nickson, Noriega-Crespo, Östlin, O'Sullivan, Ottens, Patapis, Penanen, Pietraszkiewicz, Ray, Regan, Roteliuk, Royer, Samara-Ratna, Samuelson, Sargent, Scheithauer, Schneider, Schreiber, Shaughnessy, Sheehan, Shivaei, Sloan, Tamas, Teague, Temim, Tikkanen, Tustain, van Dishoeck, Vandenbussche, Weilert, Whitehouse, \& Wolff}]{Wright2023}
Wright, G.~S., Rieke, G.~H., Glasse, A., {et~al.} 2023, Publications of the Astronomical Society of the Pacific, 135, 048003, \dodoi{10.1088/1538-3873/acbe66}

\bibitem[{Wynn-Williams {et~al.}(1974)Wynn-Williams, Becklin, \& Neugebauer}]{Wynn-Williams1974}
Wynn-Williams, C.~G., Becklin, E.~E., \& Neugebauer, G. 1974, The Astrophysical Journal, 187, 473, \dodoi{10.1086/152656}

\bibitem[{Xu {et~al.}(2004)Xu, Lees, Wang, Brown, Kleiner, \& Johns}]{Xu2004}
Xu, L.-H., Lees, R.~M., Wang, P., {et~al.} 2004, Journal of Molecular Spectroscopy, 228, 453, \dodoi{10.1016/j.jms.2004.05.017}

\bibitem[{Xu {et~al.}(2008)Xu, Fisher, Lees, Shi, Hougen, Pearson, Drouin, Blake, \& Braakman}]{Xu2008}
Xu, L.-H., Fisher, J., Lees, R., {et~al.} 2008, Journal of Molecular Spectroscopy, 251, 305, \dodoi{10.1016/j.jms.2008.03.017}

\bibitem[{Young {et~al.}(2012)Young, Becklin, Marcum, Roellig, De~Buizer, Herter, Güsten, Dunham, Temi, Andersson, Backman, Burgdorf, Caroff, Casey, Davidson, Erickson, Gehrz, Harper, Harvey, Helton, Horner, Howard, Klein, Krabbe, McLean, Meyer, Miles, Morris, Reach, Rho, Richter, Roeser, Sandell, Sankrit, Savage, Smith, Shuping, Vacca, Vaillancourt, Wolf, \& Zinnecker}]{Young2012}
Young, E.~T., Becklin, E.~E., Marcum, P.~M., {et~al.} 2012, The Astrophysical Journal, 749, L17, \dodoi{10.1088/2041-8205/749/2/L17}

\bibitem[{Zhang(2024)}]{Zhang2024}
Zhang, K. 2024, Reviews in Mineralogy and Geochemistry, 90, 27, \dodoi{10.2138/rmg.2024.90.02}

\bibitem[{Zhang {et~al.}(2013)Zhang, Tan, \& McKee}]{Zhang2013}
Zhang, Y., Tan, J.~C., \& McKee, C.~F. 2013, The Astrophysical Journal, 766, 86, \dodoi{10.1088/0004-637X/766/2/86}

\bibitem[{Zhao {et~al.}(2023)Zhao, Zhang, Wang, Qiu, Yan, Yu, Chen, \& Zou}]{Zhao2023}
Zhao, J.~Y., Zhang, J.~S., Wang, Y.~X., {et~al.} 2023, The Astrophysical Journal Supplement Series, 266, 29, \dodoi{10.3847/1538-4365/acc323}

\bibitem[{Zhu {et~al.}(2013)Zhu, Zhao, Wright, Sandell, Shi, Wu, Brogan, \& Corder}]{Zhu2013}
Zhu, L., Zhao, J.-H., Wright, M. C.~H., {et~al.} 2013, The Astrophysical Journal, 779, 51, \dodoi{10.1088/0004-637X/779/1/51}

\bibitem[{Öberg \& Bergin(2021)}]{Oberg2021}
Öberg, K.~I., \& Bergin, E.~A. 2021, Physics Reports, 893, 1, \dodoi{10.1016/j.physrep.2020.09.004}

\bibitem[{Öberg {et~al.}(2009)Öberg, Garrod, van Dishoeck, \& Linnartz}]{Oberg2009}
Öberg, K.~I., Garrod, R.~T., van Dishoeck, E.~F., \& Linnartz, H. 2009, Astronomy and Astrophysics, 504, 891, \dodoi{10.1051/0004-6361/200912559}

\bibitem[{Šimečková {et~al.}(2006)Šimečková, Jacquemart, Rothman, Gamache, \& Goldman}]{Simeckova2006}
Šimečková, M., Jacquemart, D., Rothman, L.~S., Gamache, R.~R., \& Goldman, A. 2006, Journal of Quantitative Spectroscopy and Radiative Transfer, 98, 130, \dodoi{10.1016/j.jqsrt.2005.07.003}

\end{thebibliography}
\bibliographystyle{aasjournal}

%TC:endignore

\end{document}